\documentclass[review]{elsarticle}
\usepackage[top=1in,bottom=1in,left=1in,right=1in]{geometry}
\usepackage{latexsym, graphicx, epsfig, amsmath, amssymb, amsfonts}
\usepackage{amsmath}
\usepackage{amssymb}
\usepackage{graphicx}
\usepackage{subfig}
\usepackage{tabu}
\usepackage{algorithm}
\usepackage{algpseudocode}
\usepackage{adjustbox}
\usepackage{bbold}
\usepackage{bm}
\usepackage{textcomp}
\usepackage[thinc]{esdiff}
\usepackage[dvipsnames]{xcolor}
\usepackage{comment}
\usepackage{soul}
\usepackage{multirow}
\biboptions{sort&compress}

\def\omg{{\Omega}}

\def \bb{\mathbf{b}}
\def \Tb{\mathbf{T}}
\def \tb{\mathbf{t}}

\def \ub{\mathbf{u}}

\def \sb{\mathbf{s}}
\def \xb{\mathbf{x}}

\def \yb{\mathbf{y}}

\def \rb{\mathbf{r}}

\def \real{\mathbb{R}}

\newcommand{\vertii}[1]{{\left\vert\left\vert #1
    \right\vert\right\vert}}

\newcommand\norm[1]{\left\lVert#1\right\rVert}

\begin{document}
\begin{frontmatter}

\title{Interfacing Finite Elements with Deep Neural Operators for Fast Multiscale Modeling of Mechanics Problems} 
\author{Minglang Yin\textsuperscript{ab}}
\author{Enrui Zhang\textsuperscript{c}}
\author{Yue Yu\textsuperscript{d}}
\author{George Em Karniadakis\textsuperscript{bc}\corref{cor1}}
\cortext[cor1]{Corresponding author: george\_karniadakis@brown.edu}
\address[a]{Center for Biomedical Engineering, Brown University, Providence, RI}
\address[b]{School of Engineering, Brown University, Providence, RI}
\address[c]{Division of Applied Mathematics, Brown University, Providence, RI}
\address[d]{Department of Mathematics, Lehigh University, Bethlehem, PA}

\begin{abstract}

Multiscale modeling is an effective approach for investigating multiphysics systems with largely disparate size features, where models with different resolutions or heterogeneous descriptions are coupled together for predicting the system's response. The solver with lower fidelity (coarse) is responsible for simulating domains with homogeneous features, whereas the expensive high-fidelity (fine) model describes microscopic features with refined discretization, often making the overall cost prohibitively high, especially for time-dependent problems. In this work, we explore the idea of multiscale modeling with machine learning and employ DeepONet, a neural operator, as an efficient surrogate of the expensive solver. DeepONet is trained offline using data acquired from the fine solver for learning the underlying and possibly unknown fine-scale dynamics. It is then coupled with standard PDE solvers for predicting the multiscale systems with new boundary/initial conditions in the coupling stage. The proposed framework significantly reduces the computational cost of multiscale simulations since the DeepONet inference cost is negligible, facilitating readily the incorporation of a plurality of interface conditions and coupling schemes. We present various benchmarks to assess accuracy and speedup, and in particular we develop a coupling algorithm for a time-dependent problem, and we also demonstrate coupling of a continuum model (finite element methods, FEM) with a neural operator representation of a particle system (Smoothed Particle Hydrodynamics, SPH) for a uniaxial tension problem with hyperelastic material. What makes this approach unique is that a well-trained over-parametrized DeepONet can generalize well and make predictions at a negligible cost. 


\end{abstract}

\end{frontmatter}
\textbf{keywords:} Machine Learning, Neural Operator, DeepONet, Concurrent Multiscale Coupling, Finite Element Model, Domain Decomposition

\newpage

\section{Introduction}
\label{sec:introduction}


Predicting and monitoring complex systems, where small-scale dynamics and interactions affect global behavior are ubiquitous in science and engineering \cite{weinan2011principles,alber2019integrating,dobson2010stability,tinsley2006multiscale,bazilevs2007variational}. In disciplines ranging from material fracture \cite{holian1995fracture} to design problems~\cite{fish2021mesoscopic}, models at microscale have shown their capability in representing detailed material response. However, despite their improved accuracy, the usability of microscopic models is often compromised by several computational challenges. Specifically, microscopic models require a small spatio-temporal scale to fully resolve small-scale details and capture underlying stochastic dynamics, leading to a prohibitive computational expense. Therefore, simulating material dynamics at meso- or macroscale using microscopic models is still largely beyond reach. In addition, although bottom-up approaches such as fine-grained atomistic models have provided important insights into processes at microscale, they generally do not scale up to finite-size samples~\cite{2019Wang_Concurrent,zhang2015fracture,jing2012effect}. These challenges raise the need for efficient mathematical models and algorithms, which are capable of capturing small-scale behaviors while being computationally expedient.

In the last two decades, a variety of multiscale approaches have been proposed to address these challenges. Microscopic effects often concentrate locally, whereas a continuum model can accurately describe the system in the rest of the domain. Solving such a continuum model by well-established numerical methods would reduce the computational cost. Following this domain-decomposition strategy, several works that combine continuum and microscopic models have been proposed \cite{ortiz1987method,lin2003theoretical,xiao2004bridging,nie2004continuum,tran2017automated,kevrekidis2009equation,theodoropoulos2000coarse,kevrekidis2003equation,d2021optimization} to model multiscale systems. Another approach focuses on developing a fast surrogate as the fine-scale model represented by homogenization \cite{zohdi2017homogenization,bensoussan2011asymptotic,weinan2003multiscale,efendiev2013generalized,you2022data,blumers2021multiscale}. For example, \cite{milton2002theory} considered an approximation model of a partial differential equation (PDE) that contains small-scale oscillations in its coefficients, in essence, replacing these coefficients in the model with effective properties so that the resulting solutions can adequately approximate the solutions of the original problem. However, quantifying effective properties poses a challenging task in light of the feasibility of acquiring parameter values and the level of accuracy of the homogenized PDE model compared to the original microscopic model.


Recently, deep learning algorithms have been proposed for simulating physical problems~\cite{carleo2019machine,karniadakis2021physics,zhang2018deep,cai2022physics,pfau2020ab}. In particular, neural networks have been employed in conjunction with standard numerical models to address the aforementioned challenges in multiscale modeling~\cite{wang2018multiscale,arbabi2020linking,rahman2020multiscale,peng2021multiscale,alber2019integrating,regazzoni2020machine,chattopadhyay2020data}. In~\cite{arbabi2020linking}, Arbabi et al. proposed a data-driven method that trains deep neural networks to learn coarse-scale partial differential operators based on fine-scale data. In~\cite{bhatia2021machine}, Bhatia et al. presented a novel paradigm of multiscale modeling that couples models at different scales using a dynamic-important sampling approach. A machine learning model is employed to dynamically sample in the phase space, hence enabling an automatic feedback from micro to macro scale. In~\cite{masi2021thermodynamics,masi2021thermodynamics}, Masi and his collaborators developed a thermodynamics-based artificial neural network (TANN) and applied it in multiscale modeling of materials with microstructure. Their results demonstrated that TANN is capable of implicitly learning the constitutive model from data and predicting the corresponding stress fields based on state variables. The authors also demonstrated that TANN is capable of solving boundary value problem in a multiscale system with complex microstructure using double-scale homogeneization scheme. Other applications of machine learning in multiscale modeling include parameter inference~\cite{wu2020bayesian,pled2021robust,xu2022machine,Park2021physics}, uncertainty quantification~\cite{chan2018machine,rocha2021fly}, data-driven modeling~\cite{chan2018machine,regazzoni2020machine,pyrialakos2021neural,ingolfsson2022machine,you2022data}, etc.

In the last few years, a new family of machine learning model, deep neural operators, have been proposed to learn the solution operator of a PDE system implicitly~\cite{li2020fourier,li2020neural,lu2021learning,you2022nonlocal}. Unlike another type of scientific machine learning, physics-informed neural networks (PINNs)~\cite{raissi2019physics}, these neural operators can solve a PDE system given a new instance of interface conditions or model parameters without retraining~\cite{cai2021deepm,lin2021seamless,yin2021simulating,lin2021operator,li2020multipole,goswami2022physics,mao2021deepm}. Hence, such computational advantage enables neural operators to serve as an efficient surrogate model in multiscale coupling tasks, and especially for time-depenendent multiscale problems, which even today have remained prohibitively expensive. Among the state-of-the-art neural operators, the Deep Operator Network (DeepONet) serves as a unique model with exceptional generalization capability and flexibility, which can learn the solution operator in irregular domains even in the presence of noise~\cite{lu2021comprehensive}. 
Another possibility is to use the Fourier neural operator (FNO), which is fast but is limited to complex geometries and structured data~\cite{li2020fourier,li2020neural}. 
Herein, we choose DeepONet as the surrogate model. For a more thorough comparison between DeepONet and FNO, we refer the reader to~\cite{lu2021comprehensive,kovachki2021neural}.


\begin{figure}
    \centering
	\includegraphics[width=0.9\textwidth]{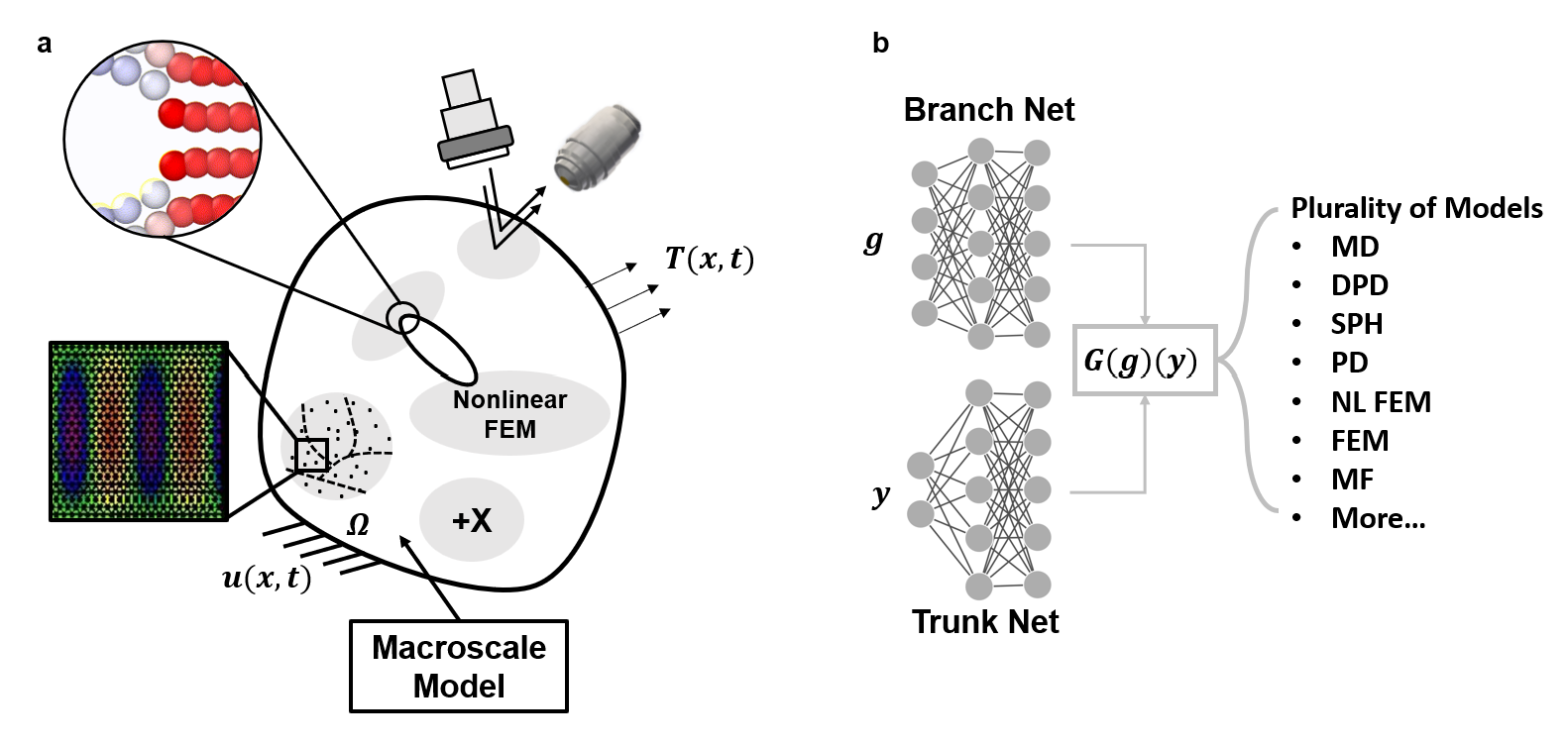} 
    \caption{\textbf{Schematic of multiscale modeling with DeepONet.} (a) In a multiscale mechanics system, traction $T(x,t)$ acts on the boundary of domain $\Omega$, which is decomposed into sub-domains described by a variety of microscopic models. A macroscopic model describes the response in the bulk region whereas nonlinear, microscopic, or data-driven models capture the detailed response in the small regions. (b) DeepONet, composed of a branch and a trunk net, is able to learn the response of the microscopic systems (from simulated or multi-modal data) and serve as a surrogate in the multiscale system. MD: molecular dynamics, DPD: dissipative particle dynamics, SPH: smoothed particle hydrodynamics, PD: peridynamics, NL FEM: nonlinear finite element, FEM: finite element, MF: multifidelity.}
    \label{fig:schmematic}
\end{figure}

The present work aims to address the aforementioned challenges by developing a new multiscale coupling framework, as demonstrated in Fig.\ref{fig:schmematic}. Specifically, the new framework couples a surrogate of the microscopic system (DeepONet) with a standard finite element method (FEM) that represents the macroscopic model. The coupling framework is flexible in choosing the domain decomposition algorithms, the types of boundary condition, or the nature of multiphysics problems (static or time-dependent). In addition, the surrogate model can learn from a plurality of fine-resolution microscopic systems or from experimental data (Fig.~\ref{fig:schmematic}(b)). Our approach is the first attempt to couple neural operators with standard numerical solvers using a concurrent coupling method.

The paper is organized as follows: In Sec. \ref{sec:deeponet}, we briefly introduce DeepONet and the adopted coupling algorithms. In Sec.~\ref{sec:results}, we report the coupling performance for a series of benchmarks, including a 2D Poisson equation, a 1D heat problem, and a uniaxial tension problem with elastoplastic materials and hyperelastic materials. We conclude by a brief discussion on the implications of the presented model in Sec.~\ref{sec:discussion}. In the appendices, we give an overview of the smoothed particle hydrodynamics (SPH) method and present additional results for FEM and SPH simulations. Finally, we provide further details on the network training and data generation.


\section{Methodology}
\label{sec:methods}

In this section, we introduce the general architecture of DeepONet (Sec.~\ref{sec:deeponet}) and the domain decomposition methods for the coupling framework (Sec~\ref{sec:coupling}). 

\subsection{Deep Operator Network (DeepONet)}
\label{sec:deeponet}

\begin{figure}
    \centering
	\includegraphics[width=0.6\textwidth]{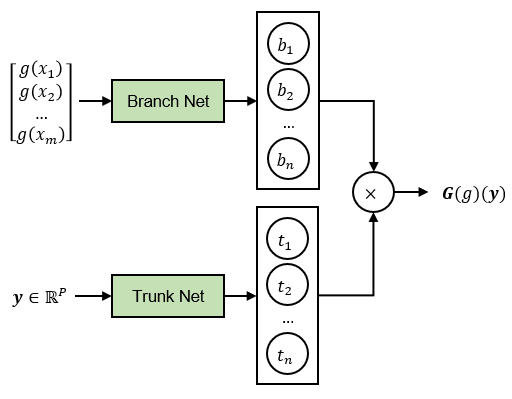} 
    \caption{\textbf{Schematic architecture of DeepONet.} DeepONet learns the mapping operator $G$ from an input function \textbf{$g$} to its corresponding output function $G(g)$. The input of branch and trunk net are $g$ and $\yb\in \real^{p}$, which represents a discretized function from $g(x_{1})$ to $g(x_{m})$ and information such as coordinates and time, respectively. Note that the output dimension of the trunk net, $n$, is consistent with that of the branch net. The final output $G(g)(\yb)$ is computed as the dot product of $\bb$ and $\tb$.
    }
    \label{fig:deeponet}
\end{figure}

We briefly summarize the general architecture of DeepONet employed in this work. Let $\Omega\subset\real^p$ be a bounded open set, which is the domain of our input and output functions; DeepONet can learn a general continuous operator between two Banach spaces of functions taking values in $\real^{d_f}$ and $\real^{d_u}$, respectively. We denote the input and output function spaces as $\mathcal{A}=\mathcal{A}(\Omega;\mathbb{R}^{d_f})$ and $\mathcal{U}=\mathcal{U}(\Omega;\mathbb{R}^{d_u})$. The network aims at approximating a mapping  $G:\mathcal{A}\rightarrow\mathcal{U}$ between an input function $g\in\mathcal{A}$ and its corresponding output function $G(g)\in\mathcal{U}$. Although DeepONet can learn mappings between vector-valued functions, for simplicity of exposition we focus on learning scalar-valued functions ($d_u=1$) in the following. For any $\yb\in\omg\subset\real^p$, $G(g)(\yb)\in \mathbb{R}$ is the evaluation of function $G(g)$ at $\yb$. As shown in Fig.~\ref{fig:deeponet}, the branch network takes a function $g$ in its discrete representation $[g(\xb_{1}), g(\xb_{2}),...,g(\xb_{m})]$ at locations $\{\xb_j\}_{j=1}^m$ as input and yields an array of operator features, $\{b_i\}_{i=1}^n$, as its output. The trunk network takes $\yb$ as input and yields another array of operator features, $\{t_i\}_{i=1}^n$. Finally, $G(g)(\yb)$ can be approximated by~\cite{lu2021learning,chen1995universal}:
\begin{equation}
\label{equ:loss}
    G(g)(\yb) \approx \sum^{n}_{i=1} b_{i}t_{i}.
\end{equation}
In this work, we adopt fully-connected neural networks as the architecture of both sub-networks. We refer the readers to~\cite{lu2021learning,lanthaler2021error} for theoretical analysis and error estimations of DeepONet. 

Regarding the loss function, we adopt the mean squared error (MSE) which measures the square of $L_{2}$ norm between model predictions and training data. Given $M$ sample input functions $g^{(i)}$, $i=1,\cdots,M$, and the ground-truth of their corresponding output functions, $G_{\text{data}}(g^{(i)})(\cdot)$ on a set of $N$ points $\{\yb_j\}_{j=1}^N$, the loss is expressed as:
\begin{equation}
\label{equ:net_loss}
    \mathcal{L} = \frac{1}{MN} \sum^{M}_{i=1}\sum^{N}_{j=1} ( G_{\text{model}}(g^{(i)})(\yb_{j}) - G_{\text{data}}(g^{(i)})(\yb_{j}) )^{2} + \hat{R},
\end{equation}
where $G_\text{model}$ is the learned operator of DeepONet for approximation and $G_\text{data}(g^{(i)})(\yb_{j}) $ denotes the data measurements for the $i$-th output function at $j$-th evaluation point. $\hat{R}$ is an additional loss term for the purpose of regularization (see Sec.~\ref{sec:poisson} and Eq.~\ref{equ:reg}).

\subsection{Coupling Methods}
\label{sec:coupling}

Here we introduce the procedure of coupling DeepONet with a numerical model. Consider a system defined on a computational domain $\omg$; we decompose the domain into $\omg_I$ and $\omg_{II}$ according to the features in each domain. These two domains are described by a macroscopic model (model I) and a microscopic model (model II), respectively. In this paper, we choose model I as FEM. Nonetheless, the proposed procedure can be generalized to couple other mesh-based models~\cite{patera1984spectral,karniadakis2005spectral,hughes2012finite,versteeg2007introduction} or particle (meshfree) models~\cite{monaghan1992smoothed,espanol1995statistical,groot1997dissipative,rapaport2004art} as shown in Fig.~\ref{fig:schmematic}(b). Model II represents a DeepONet, which serves as a surrogate for the microscopic, fine-scale model.

\begin{figure}
    \centering
	\includegraphics[width=1.0\textwidth]{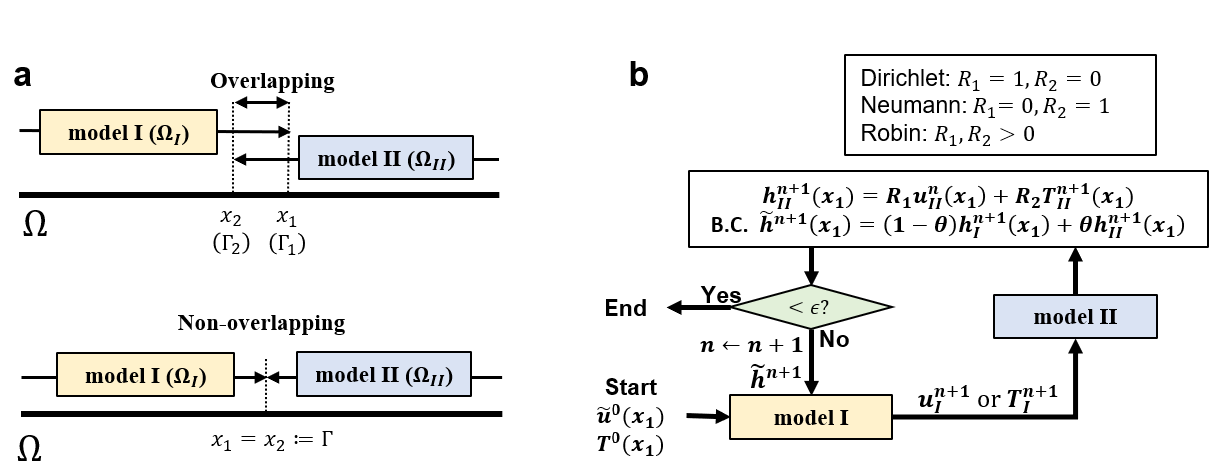} 
    \caption{\textbf{A flexible framework in domain decomposition.} The proposed coupling framework is able to adopt either (a) overlapping or non-overlapping domain decomposition coupled with Dirichlet, Neumann, or Robin boundary conditions at the interface. As an illustration, (b) presents a Robin-type boundary condition imposed on model I with interfacial solution $\hat{u}(x_{1})$ updated by a relaxation scheme. Dirichlet and Neumann boundary can be imposed by adjusting the value of $R_1$ and $R_2$. The relaxation parameter, $\theta$, is either fixed at a value or updated dynamically. $T_{II}^{n}(x_{1})$ represents the Neumann-related information at $x_{1}$ from Model II, e.g., traction or derivatives.
    }
    \label{fig:coupling}
\end{figure}

Notably, the domain decomposition can be either overlapping or non-overlapping. Fig.~\ref{fig:coupling}(a) shows an one-dimensional illustration of the domain decomposition method. For the overlapping setting, $\Gamma_1$ ($=\{x_1\}$ for this 1D case) represents the internal boundary of model I and $\Gamma_2$ ($=\{x_2\}$) is the boundary of model II.
The overlapping region is $\overline{\Omega_{I}}\cap\overline{\Omega_{II}}=[x_{2}, x_{1}]$. The two models communicate with each other by exchanging interface conditions on $\Gamma_1$ and $\Gamma_2$. For the non-overlapping setting, we have $\Gamma_1=\Gamma_2:=\Gamma$ ($x_1=x_2$ for this 1D case) and the two model exchanges interface information on $\Gamma$.
Fig.~\ref{fig:coupling}(b) and Algorithm \ref{alg:coupling} summarize the iterative procedure of the coupling framework with a Robin-type boundary condition~\cite{yu2018partitioned}. For the $n$-th iteration, we utilize quantities $(\cdot)^n$ to calculate $(\cdot)^{n+1}$. To initiate the coupling procedure ($n=0$), we start with an initial guess of the interface solution value $u^{0}(x_{1})$ and derivative-related information $T^{0}(x_{1})$ on $\Gamma_1$. Then, we take a linear combination of these quantities and forms a Robin-type boundary condition on $\Gamma_1$, namely, $\Tilde{h}^{0}(x_1)=R_{1} u^{n}(x_{1}) + R_{2}T^{0}(x_{1})$, which is applied on Model I. The method proceeds by solving for $u_{I}^{n+1}(x)$, $x\in\omg_I$, from Model I, and interpolating the computed solution or its derivatives at $\Gamma_{2}$. The interpolated information will then be transmitted to Model II as the boundary condition. Then, we solve for $u_{II}^{n+1}(x)$, $x\in\omg_{II}$, from Model II with the transmitted interface condition on $\Gamma_{2}$ and interpolate its solution on $\Gamma_{1}$. If the stopping criterion 
\begin{align}
    \vertii{u_I^{n+1}-u_I^{n}}^2_{L^2(\omg_I)}+\vertii{u_{II}^{n+1}-u_{II}^{n}}^2_{L^2(\omg_{II})}<\epsilon,
\end{align}
is satisfied, the coupling result is considered to be converged. The solution for the two domains is: 
\begin{align}
&u_I(x):=u^{n+1}_I(x),\;x\in\omg_I,\\
&u_{II}(x):=u^{n+1}_{II}(x),\;x\in\omg_{II}.
\end{align}
If the solution is not converged, we proceed to update the interface information on $\Gamma_{1}$ with a relaxation formulation: $\Tilde{h}^{n+1}(x_{1}) = (1-\theta)h_{I}^{n+1}(x_{1}) + \theta h^{n+1}_{II}(x_{1})$ where $h^{n+1}_{I, II} = R_{1} u^{n}_{I, II}(x_{1}) + R_{2}T^{0}_{I, II}(x_{1})$ are the Robin boundary condition from Model I and II. Here, the relaxation parameter $\theta \in [0, 1]$ can be either fixed or updated according to the Aitken's rule~\cite{mok2001accelerated,yu2018partitioned}.
Then, we proceed to a new iteration by transmitting the updated boundary condition $\Tilde{h}^{n+1}(x_1)$ to Model I and repeating the procedure stated above with $n\leftarrow n+1$. This procedure is repeated until the stopping criterion is satisfied. These coupling algorithms have their origin to the classical Schwarz coupling methods~\cite{lions1988schwarz,mota2017schwarz,funaro1988iterative} and an iterative patching algorithms~\cite{funaro1988iterative}.

\begin{algorithm}
\caption{Coupling Method. $\Gamma_1:=\partial\omg_I$, $\Gamma_2:=\partial\omg_{II}$. If non-overlapping, $\Gamma_1=\Gamma_2$, else $\Gamma_1\neq\Gamma_2$.}\label{alg:coupling}
\begin{algorithmic}
\State \textbf{Initialization:} Set model I with $u(x_{1}) = 0$ and model II with $u(x_{1}) = 0$ 
\State \textbf{Main Loop:} 
\For{$n=0:n_{max}-1$}
\State \textbf{Model I (FEM): } 
    \begin{itemize}
        \item Receive the interface information $h^{n}({x_{1}})$ from Model II ($x_1\in\Gamma_1$).
        \item Solve for $u^{n+1}_{I}$ from Model I.
        \item Calculate $u^{n+1}_{I}(x_{2})$ or $\frac{\partial u_{I}^{n+1}}{\partial x}|_{x_{2}}$ and pass it to Model II ($x_2\in\Gamma_2$). 
    \end{itemize}
\State \textbf{Model II (NN): }
    \begin{itemize}
        \item Receive the interface information $u^{n+1}_{I}(x_{2})$ or $\frac{\partial u_{I}^{n+1}}{\partial x}|_{x_{2}}$ from Model I ($x_2\in\Gamma_2$).
        \item Solve for $u^{n+1}_{II}/T^{n+1}_{II}$ from Model II.
        \item Calculate $h^{n+1}_{II}(x_1)=R_1 u^{n+1}_{II}(x_1)+R_2 T^{n+1}_{II}(x_1)$.
        \item Calculate $\Tilde{h}^{n+1}(x_{1}) = (1-\theta) h^{n+1}_{II}(x_{1}) + \theta h^{n+1}_{I}(x_{1})$ and pass it to Model I ($x_1\in\Gamma_1$).
    \end{itemize}
\State \textbf{If converged, stop;}
\EndFor
\end{algorithmic}
\end{algorithm}

\section{Results}
\label{sec:results}

\begin{table}
\centering
\small
\begin{adjustbox}{}
 \begin{tabular}{| c | c | c | c | c |} 
 \hline
 Problem & Model I & Model II/Training Data & Interface Condition & Overlap \\
 \hline
 2D Poisson & FEM & DeepONet/FEM & D-D, R-D & Yes  \\
 \hline
 1D Heat & FEM  & DeepONet/FEM & N-D & No  \\
 \hline
 2D Elastoplasticity & Linear Elastic FEM & DeepONet/Elastoplastic FEM & N-D & No \\
 \hline
 2D Hyperelasticity & Hyperelastic FEM & DeepONet/Hyperelastic SPH & N-D, R-D/R-N & No  \\
 \hline
\end{tabular}
\end{adjustbox}
\caption{ \textbf{Setup for the four examples.} D-D: Dirichlet-Dirichlet, R-D: Robin-Dirichlet, N-D: Neumann-Dirichlet, R-N: Robin-Neumann}
\label{table:prob_setup}
\end{table}

In this section, we present the simulation results of four benchmark problems: 2D Poisson equation, 1D heat equation, 2D elastoplasticity, and 2D hyperelasticity, to demonstrate the applicability of our method. The detailed settings of these four examples, including the choices of model I and model II, interface conditions, and whether domains overlap, are provided in Table~\ref{table:prob_setup}. For interface conditions, we studied four types of conditions, namely, Dirichlet-Dirichlet (D-D), Robin-Dirichlet (R-D), Neumann-Dirichlet (N-D), and Robin-Neumann(R-N). The first letter represents the boundary condition for Model I and the second letter for Model II. For each problem, our framework consists of two stages. In the first stage, we train a DeepONet offline to obtain a surrogate of model II. Then, we couple Model I with DeepONet in the online stage using our proposed coupling method.
Training data of elastoplastic FEM in Sec.~\ref{sec:elasto} and hyperelasticity in Sec.~\ref{sec:hyper} are generated by Abaqus~\cite{abaqus2020} and an SPH solver~\cite{ganzenmuller2015hourglass}. All the other usages of FEM (including FEM solver for model I and generation of training data for model II) are based on the FEniCS package~\cite{alnaes2015fenics} using second-order Lagrange polynomials.

\subsection{Poisson equation}
\label{sec:poisson}

\begin{figure}
    \centering
	\includegraphics[width=0.8\textwidth]{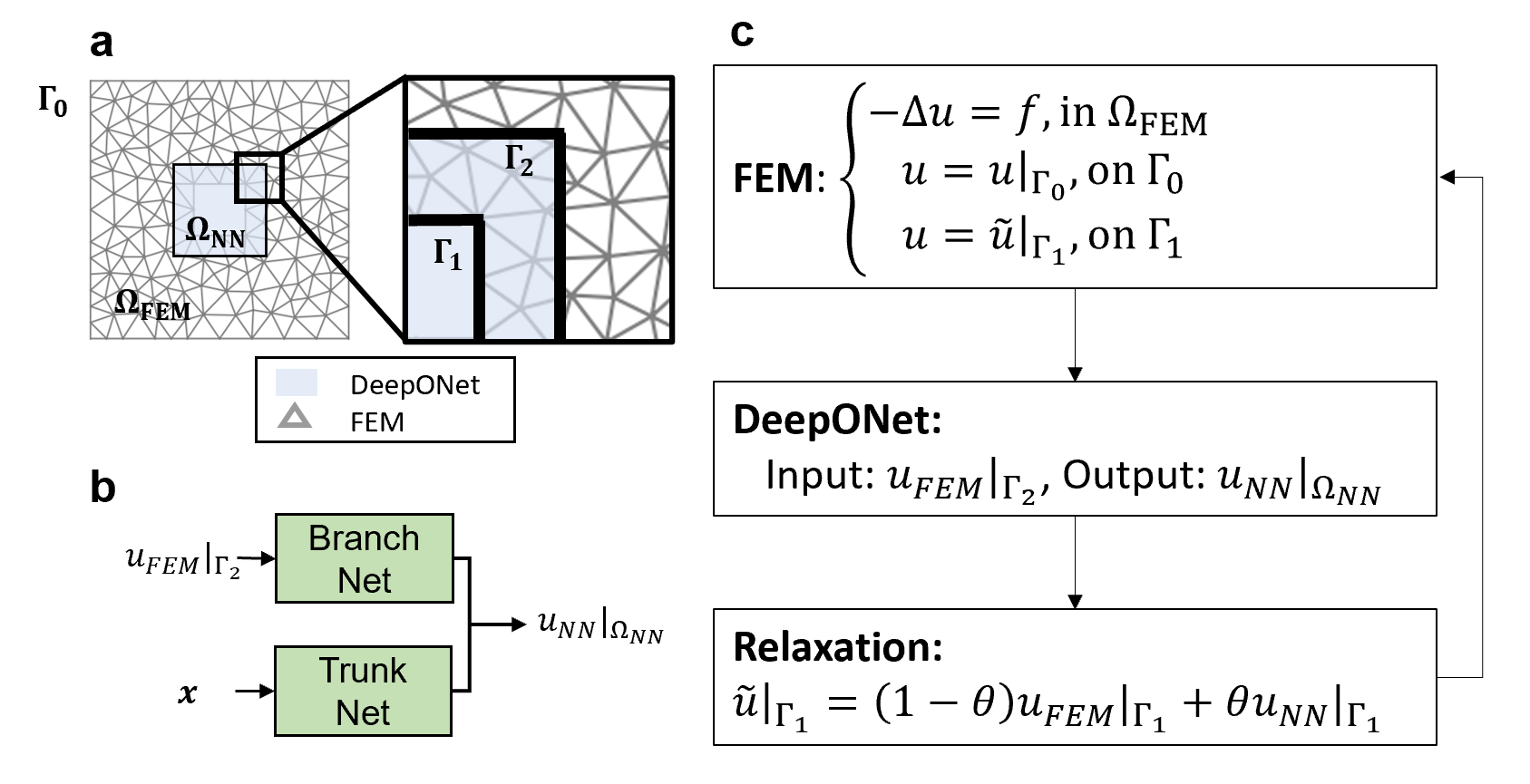} 
    \caption{\textbf{Setup of the Poisson equation.} (a) A $1\times 1$ unit square is decomposed into two overlapping regions, $\Omega_{NN}$ (light blue) and $\Omega_{FEM}$ (meshed). $\Omega_{FEM}$ is bounded by $\Gamma_{1}$ and $\Gamma_{0}$, whereas $\Omega_{NN}$ is a $0.3\times 0.3$ square in the center with boundary $\Gamma_{2}$. (b) DeepONet takes $u_{FEM}|_{\Gamma_{2}}$ as input and yields the solution in $\Omega_{NN}$, denoted as $u_{NN}|_{\Omega_{NN}}$. (c) Formulation of an overlapping D-D method. Given the boundary condition on $\Gamma_{0}$, the coupling framework iteratively updates the corresponding boundary condition on $\Gamma_{1}$. The relaxation parameter $\theta$ is either fixed or updated based on the Aitken's rule~\cite{yu2018partitioned}.}
    \label{fig:poisson}
\end{figure}

We first study the feasibility of the proposed framework for solving a static problem. Let us consider a Poisson equation described by the PDE system in $\Omega := \Omega_{FEM} \cup \Omega_{NN}$:
\begin{align}
    - \Delta u(\xb) & = f(\xb), \text{ in } \Omega = \text{[0, 1]}^{2}, \label{equ:poi_pde}\\
    u(\xb) & = u|_{\Gamma_0}(\xb), \text{ on } \Gamma_{0},\label{equ:poi_bc0}
\end{align}
where we set $f(\xb)=6$ for all $\xb\in\Omega$. As shown in Fig.~\ref{fig:poisson}(a), we decompose one domain into two overlapping subdomains, namely $\Omega_{FEM}=[0, 1]^{2} / [0.4, 0.6]^{2}$ (with an internal boundary $\Gamma_1$) and $\Omega_{NN}=[0.3, 0.7]^{2}$ (with a boundary $\Gamma_2$). In $\Omega_{FEM}$, the system is governed by Eqs.~\eqref{equ:poi_pde} and \eqref{equ:poi_bc0} with an boundary condition
\begin{align}
    u(\xb) & = \Tilde{u}|_{\Gamma_{1}}(\xb) , \text{ on } \Gamma_{1}. \label{equ:poi_bc1}
\end{align}
In this example, we first verify the efficacy of the coupling framework with a D-D interface condition. For this setup, in $\Omega_{NN}$, the system is governed by Eqs.~\eqref{equ:poi_pde} and \eqref{equ:poi_bc0} with input 
\begin{align}
    u(\xb) & = u_{FEM}|_{\Gamma_{2}} , \text{ on } \Gamma_{2}. \label{equ:poi_bc2D}
\end{align}
After presenting the results of the D-D case, we display results of parametric studies for better demonstrating the influence of other factors that may influence the convergence rate.

In the offline stage, we train a DeepONet as a surrogate of FEM for predicting solution in $\Omega_{NN}$. First, we sample a set of boundary conditions $u|_{\Gamma_{2}}$ from a random field with $\alpha=5$, a parameter in the correlation function in Eq.~\eqref{equ:grf}. Qualitatively, a smaller $\alpha$ results in a less smooth function. More details of the random field generation are provided in~\ref{sec:rand_field}. Then, we solve the Poisson equation with the randomly sampled boundary conditions using FEM, whose computational results in $\Omega_{NN}$ are collected as the training cases. We generate 1,000 cases as the training dataset of DeepONet following the aforementioned process. As shown in Fig.~\ref{fig:poisson}(b), the branch network input $u_{FEM}|_{\Gamma_{2}}$ comes from an interpolation of the FEM solution on $\Gamma_{2}$, while the trunk network takes the coordinate $\xb$ ($\in\Omega_{NN}$) as its input. The network output is $u_{NN}(\xb)$. We present more details related to the network training in~\ref{sec:net_train}.

After the completion of the offline training, we proceed to the coupling stage as shown in Fig.~\ref{fig:poisson}(c). Given a fixed boundary condition on $\Gamma_{0}$, we initiate the coupling framework with an initial guess of $u$ on $\Gamma_{1}$, which is typically zero. Then, we solve the governing equation in $\Omega_{FEM}$ with FEM and interpolate the solution on $\Gamma_{2}$ (denoted as $u_{FEM}|_{\Gamma_{2}}$), which will be used as the input of the branch network. With this input, the DeepONet then predicts $u_{NN}|_{\Omega_{NN}}$, the solution of the Poisson equation in $\Omega_{NN}$. Following that, a relaxation scheme is adopted to update the interface condition on $\Gamma_1$ as: $\Tilde{u}|_{\Gamma_{1}} = (1-\theta)u_{FEM}|_{\Gamma_{1}} + \theta u_{NN}|_{\Gamma_{1}}$, which will later be used as the boundary condition of FEM in the next iteration. The coupling iteration continues until the solution $u$ converges. 

\begin{figure}
    \centering
	\includegraphics[width=0.99\textwidth]{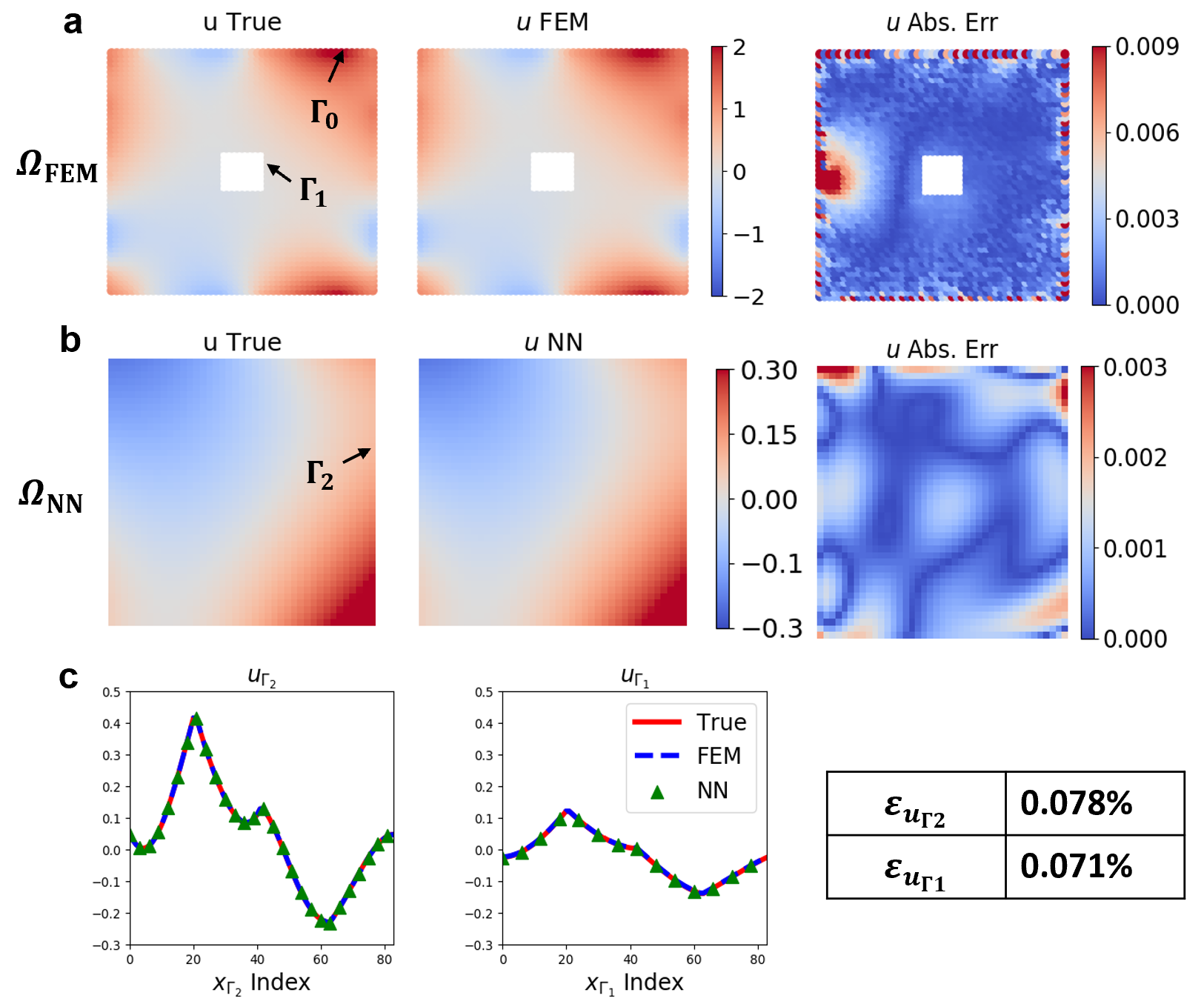} 
    \caption{\textbf{Results of coupling FEM and DeepONet for the Poisson equation.} (a-b) Model predictions from the FEM and DeepONet with the corresponding absolute errors in $\Omega_{FEM}$ and $\Omega_{NN}$. (c) Model predictions at the interfaces ($u|_{\Gamma_{1}}$ and $u|_{\Gamma_{2}}$) and the true solution (red line). The relative errors of model predictions at the interfaces are far less than 1\%. Please see Fig.~\ref{fig:poisson} for $\Omega_{NN}$ and $\Omega_{FEM}$.}
    \label{fig:poisson_comp}
\end{figure}

We show the performance of our coupling framework for the Poisson equation in Fig.~\ref{fig:poisson_comp}. We fix the relaxation parameter as $\theta=0.5$ and impose a boundary condition on $\Gamma_0$ from the testing dataset (unseen to the DeepONet in the training stage).  Figs.~\ref{fig:poisson_comp}(a-b) show a comparison between the FEM ground truth (first column; computed directly in $\Omega$) and the predictions from the coupling framework (second column; first row in $\Omega_{FEM}$ from FEM, second row in $\Omega_{NN}$ from DeepONet) together with their difference (third column). We observe an agreement between the coupling predictions and the ground truth. The maximum absolute error presented in the third column is less than 1\%. We also note that the error is larger in $\Omega_{FEM}$, especially in the region close to its external boundary $\Gamma_{0}$. The observed error is mostly dominated by the interpolation accuracy in the FEM, not the error of the coupling framework. We further show a quantitative comparison between models prediction and ground-truth on $\Gamma_{1}$ and $\Gamma_{2}$ in Fig.~\ref{fig:poisson_comp}(c): the predictions from coupled FEM/NN (blue dashed lines and green triangles, respectively) accurately reproduce the true solution (red lines) with a relative error at around 0.07\%. 

\begin{figure}
    \centering
	\includegraphics[width=1.0\textwidth]{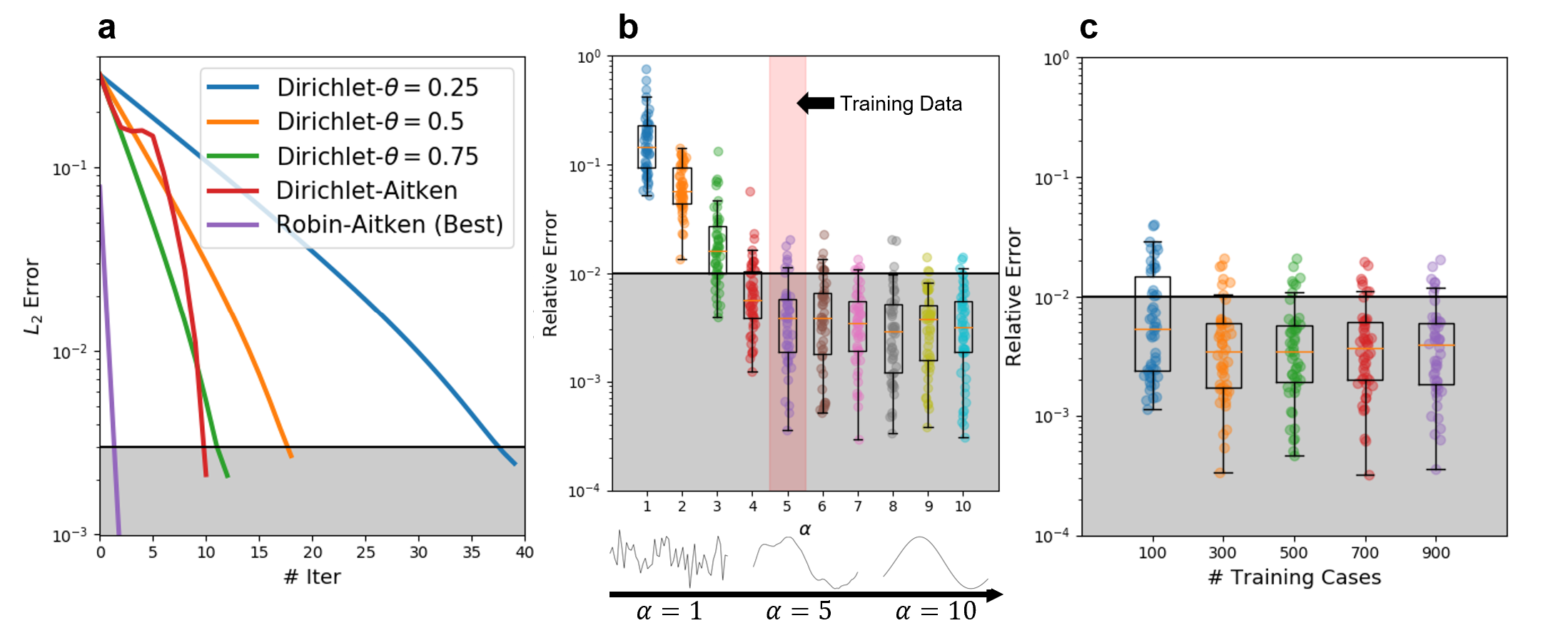}
    \caption{\textbf{Parametric studies of the Poisson equation.} (a) Given a boundary condition on $\Gamma_{0}$, the convergence history of the coupling model varies with the coupling method (Robin or Dirichlet) and parameter $\theta$. (b) The model is trained with data generated from a random field with $\alpha=5$. We test the generalization ability represented by the relative errors of the coupling model for cases corresponding to $\alpha=1$ to $10$. (c) Box plots of relative errors for DeepONets trained with $100$ to $900$ cases. Each column shows the coupling results by testing with 50 different cases. The shaded area indicates that the relative error on $\Gamma_{1}$ is less than 1\%. Notice that we present the error with respect to the ground truth to show the convergence and accuracy of our framework.}
    \label{fig:poisson_converg}
\end{figure}


In addition to the results shown in Fig.~\ref{fig:poisson_comp}, we also conducted parametric studies on diverse factors that influence the performance of our framework, including: convergence of the framework with different coupling methods (Fig.~\ref{fig:poisson_converg}(a)); the coupling accuracy influenced by the extrapolation capability of the DeepONet (Fig.~\ref{fig:poisson_converg}(b)) and the number of training cases (Fig.~\ref{fig:poisson_converg}(c)). In Fig.~\ref{fig:poisson_converg}(a), convergence of errors with different boundary conditions is plotted against iterations. The shaded area indicates that the relative error is less than 1\%. Specifically, given various values of the relaxation parameter $\theta$ ($\theta=0.25, 0.5$ and $0.75$), the displacement errors with Dirichlet boundary conditions satisfy the stopping criterion ($L_{2}$ error less than $2\times 10^{-3}$, or equivalently, relative error less than 1\%) at iteration 37, 18, and 12, respectively. With the Aitken's relaxation strategy for $\theta$ (see, e.g.,~\cite{yu2018partitioned}), the coupling error (denoted as ``Dirichlet-Aitken'') converges a bit faster than that of a fixed relaxation parameter $\theta=0.75$. We also adopt a Robin-type boundary (Robin-Aitken) with dynamic update in $\theta$ (purple line). In this case of Robin boundary condition, the system in domain $\Omega_{FEM}$ is governed by Eqs.~\eqref{equ:poi_pde} and \eqref{equ:poi_bc0} with a Robin boundary:
\begin{align}
    R_{1}u + R_{2}\frac{\partial u}{\partial n} &= g, \text{ on $\Gamma_{1}$},
\end{align}
where we set $R_{1}=1$ and $R_{2}=1$. Since the FEM needs both the information of the solution and its derivative in the normal direction from DeepONet to update the Robin boundary condition, we adjust the training loss of the network following the strategy. The regularization term $\hat{R}$ in Eq.~\eqref{equ:net_loss} is set as:
\begin{equation}
    \label{equ:reg}
    \hat{R} = \left(\left.\frac{\partial u}{\partial n}\right|_{NN} - \left.\frac{\partial u}{\partial n}\right|_{true}\right)^{2}, \xb\in \Gamma_{1}
\end{equation}
to regularize the partial derivative with respect to the normal direction of $\Gamma_{1}$. In Eq.~\eqref{equ:net_loss}, the partial derivative term is computed by taking the automatic differentiation of the network output with respect to the trunk net input $\xb$~\cite{goswami2022physics,wang2021learning,yin2021non,zhang2020physics,zhang2022void}. With the employment of the Robin boundary condition, we observe that the method only takes two iterations to reach a relatively small error. 

In Fig.~\ref{fig:poisson_converg}(b), we show the generalization ability of DeepONet and its impact on the accuracy of the coupling framework by testing with boundary conditions outside the training region. The minimal relative errors of $u|_{\Gamma_{2}}$ are plotted against the correlation length $\alpha$ of the random field that was used to generate training data. When the correlation length $\alpha$ increases, the sampled curves become smoother and vice versa. Notice that we train the network on training samples with $\alpha=5$ and test its performance on $\alpha$ ranging from 1 to 10, each with 50 testing cases. The relative errors drop from around 10\% to less than 1\% with $\alpha$ increasing from 1 to 5. For $\alpha >5$, the errors are statistically stable.

Fig.~\ref{fig:poisson_converg}(c) exhibits the accuracy of the coupling framework as a function of the generalization ability of DeepONet, which is reflected by the number of training cases for DeepONet. The median of the relative errors of $u|_{\Gamma_{1}}$ decreases slightly when the number of training cases increases from 100 to 300 and then stays statistically stable even with further increase. The computational results implicitly suggest that after 300 training cases, the errors are most contributed by the accuracy of boundary interpolation, not the generalization of DeepONet. The shaded area in (b-c) denotes relative error at 1\%.

\subsection{Heat equation}
\label{sec:heat}

\begin{figure}
    \centering
	\includegraphics[width=1.0\textwidth]{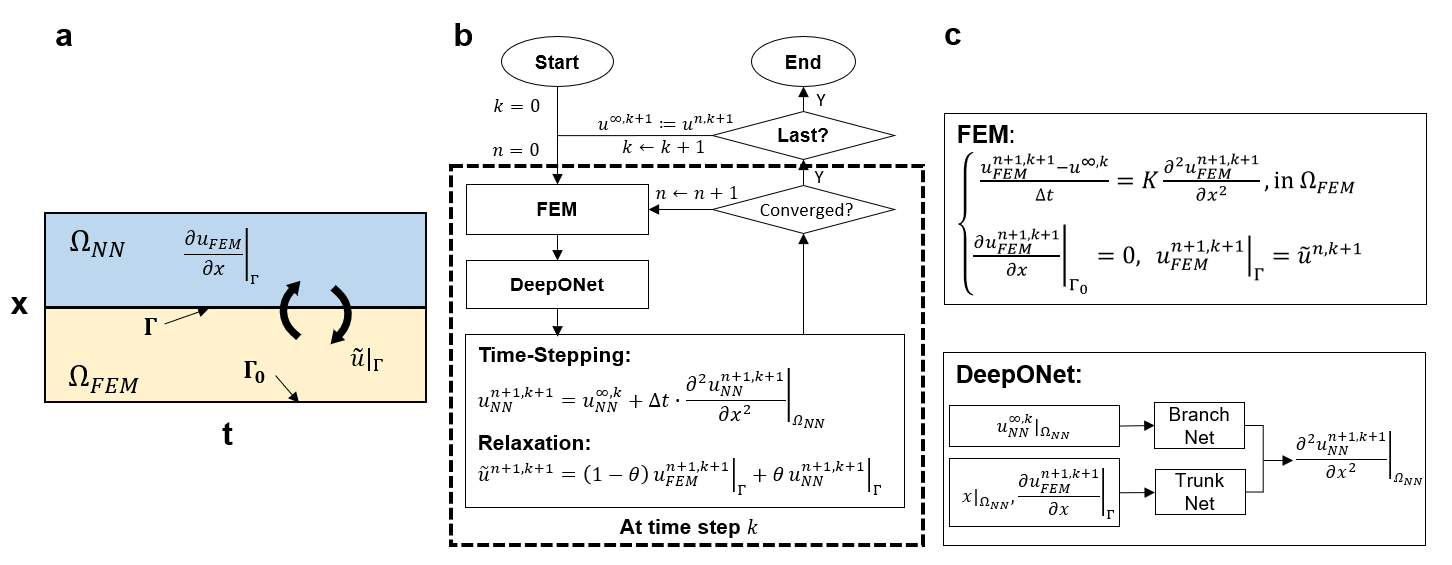} 
    \caption{\textbf{Setup of the time-dependent problem (heat equation).} (a) The spatio-temporal domain is decomposed into two sub-domains with $\Gamma$ as the interface between FEM and NN domain. (b-c) Formulation of the coupling framework for the heat equation. The framework first solves for $u^{n+1,k+1}_{FEM}$ from FEM with updated information at the interface $\Tilde{u}$ at $k$-th step. Then, the computed derivative at $\Gamma$ is transmitted into DeepONet, which predicts the spatial derivative in $\Omega_{NN}$ (c). The solution in $\Omega_{NN}$ is computed based on a time-stepping scheme, followed by a relaxation update (b). The initialization at $k=0$ and $n=0$ is described in the main text.} 
    \label{fig:heat_setup}
\end{figure}

Next, we investigate the performance of the coupling framework for a dynamic problem, namely, 1D heat equation, with a N-D method. Consider a PDE system in $\Omega := \Omega_{FEM}\cup \Omega_{NN}$:
\begin{align}
    \frac{\partial u}{\partial t}  = K \frac{\partial^{2} u}{\partial x^{2}}, &\quad\text{ for } (x, t) \in \Omega := {[0, 1.0]}\times {[0, 10.0]},  \\
    u(x,0)  = u_{0}(x), &\quad\text{ for } x \in {[0, 1.0]}, \\
    u_{x}(x,t)  = 0, &\quad\text{ for }(x,t) \in \Gamma_{0} := \{0,1.0\} \times [0, 10.0],
\end{align}
where the thermal diffusivity $K$ is set as 0.1. As shown in Fig.~\ref{fig:heat_setup}(a), we decompose the computational domain $\Omega$ into two non-overlapping subdomains: $\Omega_{FEM} := (x,t) \in [0, 0.5]\times [0, 10.0]$ for FEM and $\Omega_{NN} := (x,t) \in [0.5, 1]\times [0, 10.0]$ for DeepONet. The interface is $\Gamma:=\{0.5\}\times [0,10]$. For the FEM subdomain $\Omega_{FEM}$, we set a Dirichlet boundary condition at the interface $\Gamma$:
\begin{align}
    u(x,t) = \Tilde{u}(x,t) \quad \text{ for } (x,t) \in \Gamma,
\end{align}
where $\Tilde{u}$ is the updated and relaxed solution at the interface. For the DeepONet subdomain $\Omega_{NN}$, we impose the interfacial flux from FEM $K \frac{\partial u_{FEM}}{\partial x}$ as input.

The coupling method is illustrated in Fig.~\ref{fig:heat_setup}(b-c). The solution process is a nested loop with indices $n$ and $k$, where $k$ represents the time step and $n$ is the current iteration step. To initiate the framework, we set both $n$ and $k$ as zero with an initial guess of interfacial flux for the FEM. The FEM solves for the solution ($u^{n+1,k+1}_{FEM}|_{\Omega_{FEM}}$) in $\Omega_{FEM}$ at time step $k+1$ given the solution at the previous time step $k$, denoted as $u_{FEM}^{\infty,k}|_{\Omega_{FEM}}$. Then, we transmit the flux at $\Gamma$, $K\frac{\partial u^{n+1,k+1}_{FEM}}{\partial x}|_{\Gamma}$, to DeepONet as a part of the trunk net input (Fig.~\ref{fig:heat_setup}(c)). The branch network takes the system solution in $\Omega_{NN}$ at $k$-th time step as input. The output of DeepONet estimates an approximation of the diffusion term $K\frac{\partial^{2} u^{n+1,k+1}_{NN}}{\partial x^{2}}$. The solution in $\Omega_{NN}$ is then calculated by a semi-discretized heat equation with the backward Euler method: $u_{NN}^{n+1,k+1}(x)=u_{NN}^{n,k}(x)+\Delta t \cdot K\frac{\partial^{2} u^{n+1,k+1}_{NN}}{\partial x^{2}}$. Next, the interfacial flux $\Tilde{u}^{n+1,k+1}$ is updated by the relaxation scheme:
\begin{equation}
    \Tilde{u}^{n+1,k+1} = \left.(1-\theta)u^{n+1,k+1}_{FEM}\right|_{\Gamma_{1}} + \left.\theta u^{n+1,k+1}_{NN}\right|_{\Gamma_{1}}
\end{equation}
In this example, we fix the relaxation parameter $\theta=0.5$. If the updated interfacial flux is not yet converged, we continue to the next iteration. Otherwise, if the flux is converged, then we proceed to the next time step in the outer loop and restart the inner loop with the reset iteration step $n=0$.

Herein, we summarize the training procedure of DeepONet. We generate 1,000 initial conditions in $x\in[0.5, 1.0]$ from a Gaussian random field with constant mean $0$ and correlation length 0.3. For each case, we randomly sample the flux at $u|_{\Gamma_1}$ 20 times based on a uniform distribution from -3 to 3 as the boundary conditions. The training data of DeepONet is then generated by running FEM simulations on $\Omega_{NN}$ with the sampled initial/boundary conditions. Note that training a DeepONet in a spatio-temporal domain with disparate boundary/initial condition is a data-demanding task: one needs to sample in the spatio-temporal domain. Also, performance of the framework may be deteriorated when the network extrapolates the solution outside the training domain. Hence, we alleviate the challenges by training the network to learn the implicit spatial derivative operator $K\frac{\partial^{2} u}{\partial x^{2}}$ with an explicit method to advance in time. In practice, we acquire the spatial derivative $K\frac{\partial^{2} u}{\partial x^{2}}$ in $\Omega_{NN}$ from FEM simulations running from from $t=0$ to $1.0$ with $dt=0.1$. The simulated spatial derivative is then collected and utilized for DeepONet training. For more details of the network training, we refer the reader to Sec.~\ref{sec:net_train}.

\begin{figure}
    \centering
 	\includegraphics[width=1.0\textwidth]{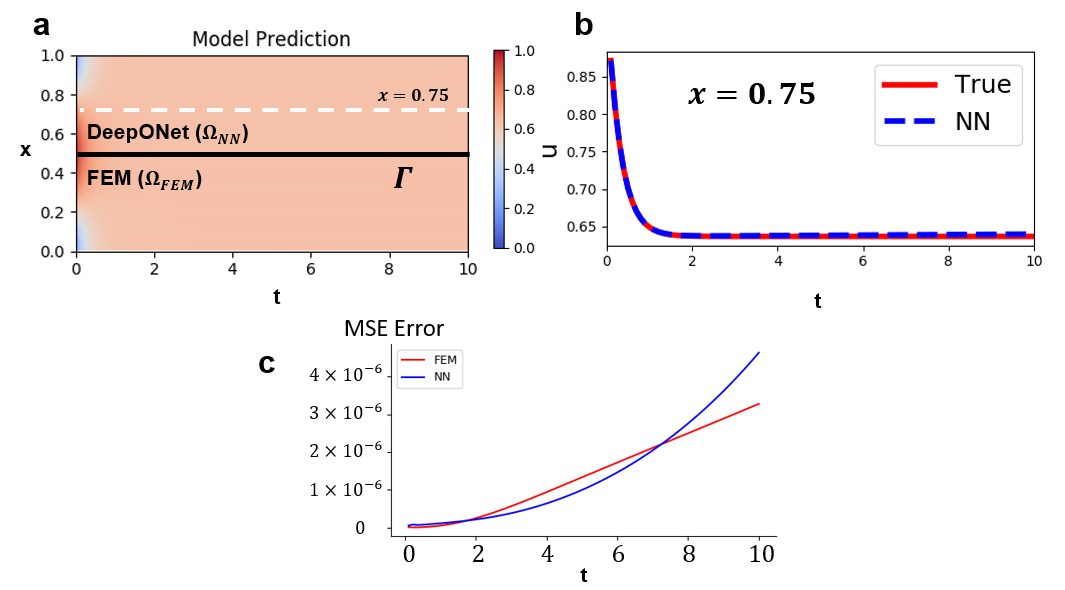} 
    \caption{\textbf{Results of coupling FEM and DeepONet for the heat equation.} (a) Model predictions from DeepONet and FEM. DeepONet predicts the solution in $x\in \Omega_{NN}$ while FEM predicts the solution for $x\in \Omega_{FEM}$. The black line at $x=0.5$ denotes the interface of the two domains. The solution at $x=0.75$ (indicated by the white dashed line) is presented in (b). (c) The mean square errors of FEM (red) and DeepONet (blue) vs time (t).}
    \label{fig:heat}
\end{figure}

The coupling results are shown in Fig.~\ref{fig:heat}. Fig.~\ref{fig:heat}(a) shows the prediction from both models in the spatio-temporal domain $\Omega$. In Fig.~\ref{fig:heat}(b), the solution of DeepONet at $x=0.75$ is plotted against time. The prediction from DeepONet shows a good agreement with the ground truth solution even for the temporal region outside the training dataset ($t > 1$), indicating that the proposed framework works well for extrapolation. 
The mean-squared errors (MSE) of both models in the coupling framework are plotted against time in Fig.~\ref{fig:heat}(c): the coupling framework shows stability and accuracy over long-time integration. Although the prediction errors grow with the increase of $t$, we note that the errors accumulate at a relatively low rate, demonstrating that the proposed coupling framework is capable of solving a time-dependent system.

\subsection{Elastoplasticity}
\label{sec:elasto}

\begin{figure}
    \centering
	\includegraphics[width=1.0\textwidth]{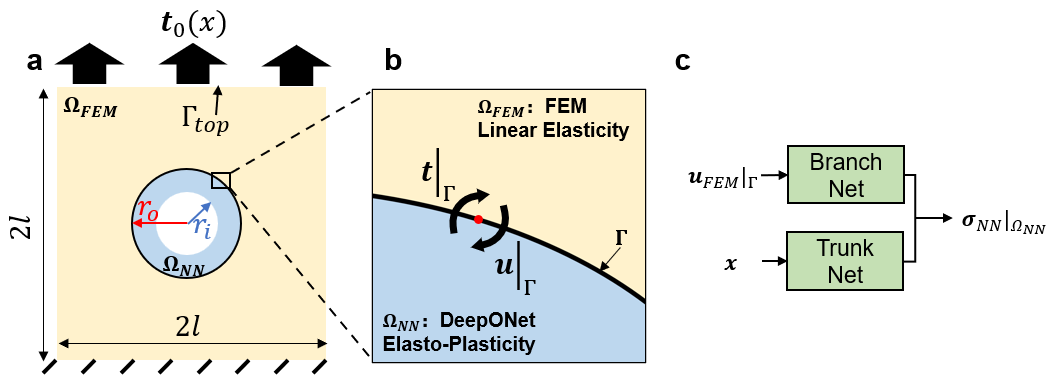} 
    \caption{\textbf{Setup of the elastoplasticity problem.} (a) A $2l\times 2l$ solid plate is clamped on the bottom edge with a traction $\mathbf{t}_{0}$ distributed on the top edge. (b) The yellow region ($\Omega_{FEM}$) is modeled with linear elasticity, whereas the blue region ($\Omega_{NN}$) is modeled with elastoplasticity. Interfacial displacement (Dirichlet boundary condition) $\ub|_{\Gamma}$ computed from FEM is transmitted to DeepONet as input. (c) The DeepONet estimates the stress components in $\Omega_{NN}$, $\sigma_{NN}|_{\Omega_{NN}}$, based on which the interfacial traction $t_{NN}|_{\Gamma}$ can be calculated accordingly. The predicted traction is updated and provided to FEM as a Neumann-type boundary condition. In this example, we set the interior circle radius $r_{i}=0.1$, the interface circle radius $r_{o}=0.3$, and the plate size $l=1$.}
    \label{fig:elastoplasticity}
\end{figure}

In this section, we test the proposed framework for predicting the elastoplastic behavior of a solid material. As shown in Fig.~\ref{fig:elastoplasticity}(a), we consider a plane strain problem for a square-shaped solid of size $2l\times 2l$ with a circular void of radius $r_i$, where vertical tension is applied on its top edge. In this example, we take $l=1$, $r_{i}=0.1$. For linear elastic materials, the kinematics, constitutive relation, and equilibrium equations are as follows: 
\begin{align}
    \bm{\varepsilon} & = \frac{1}{2}(\bm{\nabla} \ub + \bm{\nabla} \ub^{\text{T}}),\\
    \bm{\sigma} &= \lambda \text{tr}(\bm{\varepsilon})\mathbf{I} + 2\mu\bm{\varepsilon}, \\
    \mathbf{0} &= \bm{\nabla} \cdot \bm{\sigma}+\bb ,\label{eqn:EP-E-equil}
\end{align}
where $\bm{\varepsilon}$,  $\ub$, and $\bm{\sigma}$ are the strain, displacement, and (Cauchy) stress; $\lambda$ and $\mu$ are Lam\'{e} moduli, which we take as $\lambda=0.5769$ and $\mu=0.3846$, and $\mathbf{I}$ is the identity tensor. In this example, we consider a solid subject to no body load, and therefore set the body force term $\bb$ as zero. 

To model the plastic behavior of the material, we consider small-deformation, rate-independent elastoplasticity with isotropic hardening. The additive decomposition of the strain tensor writes  $\bm{\varepsilon} = \bm{\varepsilon}^\text{e} + \bm{\varepsilon}^\text{p}$, where $\bm{\varepsilon}^\text{e}$ and $\bm{\varepsilon}^\text{p}$ are the elastic and plastic strains, respectively. The elastic strain $\bm{\varepsilon}^\text{e}$ is related to the stress by
\begin{equation}\label{eqn:plasticity}
\bm{\sigma} = \lambda \text{tr}(\bm{\varepsilon}^{\text{e}})\mathbf{I} + 2\mu\bm{\varepsilon}^{\text{e}}.
\end{equation}
The plastic strain $\bm{\varepsilon}^\text{p}$ is purely deviatoric (i.e., $\text{tr}(\bm{\varepsilon}^\text{p})=0$). We define the deviatoric stress $\sb$, the increment of the equivalent plastic strain $\text{d}\bar{\varepsilon}^\text{p}$, and the equivalent tensile stress (Mises stress) $\bar{\sigma}$ as
\begin{align}
    \sb&=\bm{\sigma}-\frac{1}{3}\text{tr}(\bm{\sigma})\bm{I},\\
    \text{d}\bar{\varepsilon}^\text{p}&=\sqrt{\frac{2}{3}\bm{\text{d}\varepsilon^\text{p}}:\text{d}\bm{\varepsilon^\text{p}}},\\
    \bar{\sigma}&=\sqrt{\frac{3}{2}\sb:\sb},
\end{align}
respectively. The flow direction $\bm{N}^\text{p}$ and the increment of the plastic strain d$\bm{\varepsilon}^\text{p}$ are given by
\begin{align}
    \bm{N}^\text{p}&=\sqrt{\frac{3}{2}}\frac{\sb}{\bar{\sigma}},\\
    \text{d}\bm{\varepsilon}^\text{p}&=\sqrt{\frac{3}{2}}\text{d}\bar{\varepsilon}^\text{p}\bm{N}^\text{p}.
\end{align}
Then, the yield function $f$ can be defined as
\begin{equation}
    f=\bar{\sigma}-Y(\bar{\varepsilon}^\text{p}),
\end{equation}
where the linear strain-hardening function $Y$ is taken as
\begin{equation}
    Y(\bar{\varepsilon}^\text{p})=Y_0+H_0\bar{\varepsilon}^\text{p},
\end{equation}
The initial strength $Y_0=0.1$ and the hardening modulus $H_0=0.3$ are taken as two constant material parameters. The aforementioned mechanical quantities are subject to the Kuhn-Tucker complementary conditions:
\begin{equation}
    f\leq 0, \quad\text{d}\bar{\varepsilon}^\text{p}\geq 0,\quad(\text{d}\bar{\varepsilon}^\text{p})f=0.
\end{equation}
In addition, when $f=0$, the consistency condition $\text{d}\bar{\varepsilon}^\text{p}\text{d}f=0$ also needs to be satisfied.

Due to the setup of our boundary value problem, the plastic deformation concentrates around the void whereas the material is dominated by elastic behaviors in regions away from the void. Hence, we decompose the square domain into two non-overlapping subdomains (see Fig.~\ref{fig:elastoplasticity}(a-b)): the internal region $\Omega_{NN}$, which is an annulus with interal radius $r_i=0.1$ and external radius $r_o=0.3$ (on $\Gamma$), modeled by DeepONet as a surrogate for the solid's elastoplastic response; the external region $\Omega_{FEM}$, modeled by FEM for linear elasticity. The two regions share a common interface on $\Gamma$. In $\Omega_{NN}$, we train several DeepONets as surrogates of each stress components to capture the plastic behavior. First, we sample 1,000 displacement boundary conditions on the top edge using the sampling method described in~\ref{subsec:rand_ep}, and employ the sampled data as boundary conditions. Then, we solve for the displacement and stress fields in the entire domain with a FEM solver based on the elastoplasticity model described above.
Next, we collect the simulation results in $\Omega_{NN}$, which will be employed as the training data of DeepONets. As depicted in Fig.~\ref{fig:elastoplasticity}(c), DeepONet can predict the corresponding Cauchy stresses in $\Omega_{NN}$ with input as the displacement at the interface. In 2D problems, the Cauchy stress $\bm{\sigma}(\xb)$ for each material point is a $2\times 2$ symmetric matrix. Therefore, to model the stress we only need to predict its three components, namely, $\sigma_{11}$, $\sigma_{12}$, and $\sigma_{22}$. For each of these components, we train an independent DeepONet separately as a surrogate of these quantities. The traction at the interface $\mathbf{t}|_{\Gamma}$ is calculated accordingly based on the predicted stress from the DeepONets. More details regarding the training of this network are presented  in~\ref{sec:net_train}.

Then, we employ the trained DeepONets in the coupling framework. As depicted in Fig.~\ref{fig:elastoplasticity}, DeepONets and FEM communicate at the interface $\Gamma$ by transmitting the information of displacement and traction (Fig.~\ref{fig:elastoplasticity}). The interfacial displacement, $\ub_{FEM}|_{\Gamma}$, is computed in FEM and transmitted to the DeepONets as the input of the branch network. Then, the network solves for the corresponding stress in $\Omega_{NN}$ and calculates the traction at the interface ($\tb_{NN}|_{\Gamma}$). In the next iteration, the computed traction $\tb_{NN}|_{\Gamma}$ will be imposed as the boundary condition of the FEM model. In this example, we employ a relaxation scheme for the traction from the DeepONets. The relaxation parameter $\theta$ is fixed at 0.5. 

\begin{figure}
    \centering
	\includegraphics[width=1.0\textwidth]{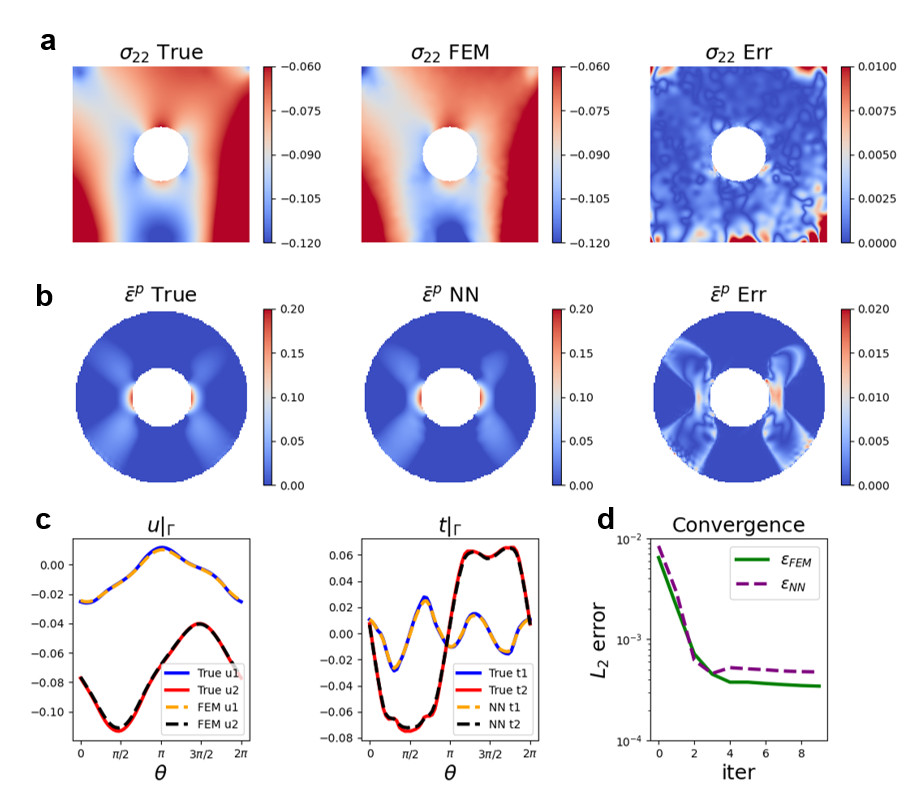} 
    \caption{\textbf{Results of coupling FEM and DeepONet in elastoplaticity.} (a-b) Results of the normal stress in $y$ direction ($\sigma_{22}$) and equivalent plastic strain ($\bar{\varepsilon}^\text{p}$). From left to right: True value, FEM/DeepONet predicted value from coupling, and their absolute errors. (c) Displacement and traction at the interface $\Gamma$. True values and predicted values from coupling are presented. Solid lines: displacement and traction of true data in x (blue) and y (red) directions. Dashed lines: displacement and traction of the model predictions in x (yellow) and y (black) directions. (d) The history of the relative errors of the coupling model at the interface. $\varepsilon_\text{FEM}$ and $\varepsilon_\text{NN}$ refer to the $L_{2}$ error of displacement and traction from FEM and DeepONet, respectively. Notice that we present the error with respect to the ground truth to show the convergence and accuracy of our framework.} 
    \label{fig:elastoplastic_comp}
\end{figure}

We present the coupling results with a N-D interface condition in Fig.~\ref{fig:elastoplastic_comp}. In Figs.~\ref{fig:elastoplastic_comp}(a-b), we plot the ground truth solution from FEM in the first column, the predictions from our coupling framework in the second column, together with their differences in the third column. In plot (a), we show the results of the normal stress ($\sigma_{22}$) in the vertical direction of $\Omega_{FEM}$. The results of the equivalent plastic strain, $\bar{\varepsilon}^\text{p}$ in $\Omega_{NN}$ are provided in plot (b). Although the coupling framework has generally well reproduced the stress component $\sigma_{22}$ in the bulk region of $\Omega_{FEM}$, there exist relatively large errors at the top edge and the bottom corners. These errors either originate from numerical interpolation in the FEM solver or are caused by a reduced solution regularity. Apart from these regions, the errors are controlled at a low value with the maximum relative error lower than 10\%. In $\Omega_{NN}$, the profile of $\bar{\varepsilon}^\text{p}$ is well captured by the coupling scheme with the maximum relative error of $\bar{\varepsilon}^\text{p}$ less than 10\%. The results demonstrate that plasticity in the region of interest is well predicted by the surrogate model. To provide a further quantitative verification of our coupling framework, we plot the ground-truth, the predictions of displacement, and traction components on $\Gamma$ in Fig.~\ref{fig:elastoplastic_comp}(c) as functions of $\theta$, the angle in polar coordinate. The solid lines denote the ground-truth results from the nonlinear FEM. The dashed lines denote the predicted values of the framework. The model predictions match the ground truth well, with relative errors smaller than 2\%. Fig.~\ref{fig:elastoplastic_comp}(d) shows the efficiency of the coupling framework: the $L_{2}$ errors between model predictions and the ground truth reach a plateau at the fourth iteration. 

\subsection{Hyperelasticity}
\label{sec:hyper}

In the previous examples, DeepONet was trained based on data generated from FEM and was coupled with another FEM in the online stage. These examples illustrated the capability of our coupling framework in solving both static and dynamic problems. In this section, we demonstrate the capability of our framework in concurrently coupling a continuum model (FEM) with a surrogate from smooth particle dynamics (SPH) for describing a microscopic particle system. We consider using the coupled framework to predict the mechanics of a hyperelastic material. We first derive the energy minimization formulation of the continuum model. We denote by $\psi$ the strain energy density of the hyperelastic model and seek to find a displacement field $\ub: \Omega \rightarrow \mathbb{R}^{2}$ that minimizes the total potential energy $\Psi$:
\begin{align}
    \Psi & = \int_{\Omega}\psi(\ub)\text{d}\xb - \int_{\Omega}\bb\cdot \ub \text{d}\xb - \int_{\Gamma_{N}} \Tb_0\cdot \ub \text{d}s.
\end{align}
Here, $\bb$ denotes the body force in $\omg$, $\Tb$ is the traction load applied on the Neumann boundary $\Gamma_N$. Hence, the total potential energy $\Psi$ is the integration of strain energy density, $\psi$, over the entire domain $\Omega$, deduced by the energy contributions from the body force $\bb$, the traction $\Tb$. 

\begin{table}
\centering
\begin{adjustbox}{}
 \begin{tabular}{|c | c cccc|} 
 \hline
 Parameter &  $\mu$ & $K$ & $k_1$& $k_2$&$\alpha$ \\
 \hline
 Value & 0.3846 & 0.8333& 0.1& 1.5& $\pi$/2\\
 \hline
\end{tabular}
\end{adjustbox}
\caption{ \textbf{Parameters value of the HGO model.} We set the value of $k_{1}, k_{2}$, and $\alpha$ the same for $i = 1$ and $2$. }
\label{table:hyper_param}
\end{table}

\begin{figure}
    \centering
	\includegraphics[width=1.0\textwidth]{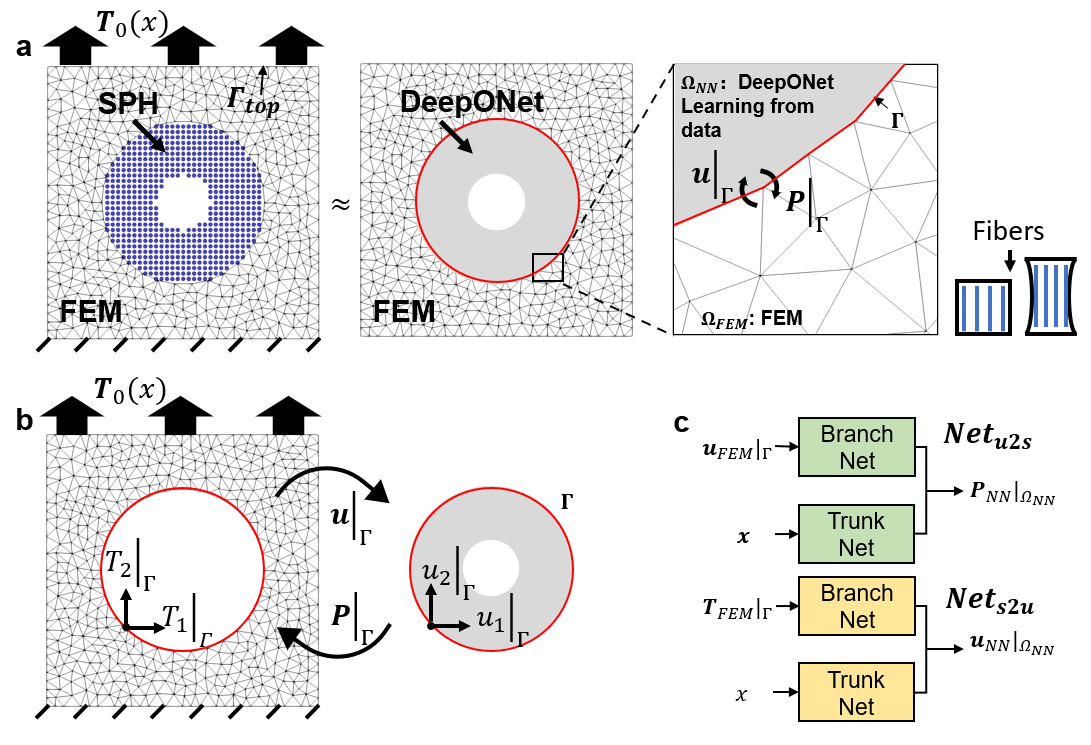} 
    \caption{\textbf{Setup of the hyperelasticity problem.} (a) A unit square is decomposed into $\Omega_{NN}$ and $\Omega_{FEM}$. We impose a non-uniform traction boundary condition on the top edge and fix the displacement at the bottom and train multiple DeepONets to represent the mechanics of an SPH model. The material is reinforced by fibers in the vertical direction. Information at the interface (displacement $\ub$ and first Piola-Kirchhoff stress $\mathbf{P}$) is transmitted between DeepONet and FEM. (b) Traction and displacement in different directions are exchanged at the interface $\Gamma$. FEM predicts the external domain while the DeepONet is trained based on SPH data. (c) Two types of DeepONet are proposed to predict the mechanics of the system: predicting stresses based on displacement information (green boxes) and vice versa (yellow boxes).}
    \label{fig:hyper_schematic}
\end{figure}

We consider the Holzapfel-Gasser-Odgen (HGO) model~\cite{holzapfel2000new} to describe the constitutive behavior of the material in this example. Essentially, the material is hyperelastic, anisotropic, fiber-reinforced in diverse directions. Its strain energy density is:
\begin{align}
    \psi & = \frac{\mu}{2}(I_{1} - 3) - \mu\ln(J) + \frac{k_{1}}{2k_{2}}\sum^{2}_{i=1}(\exp{(k_{2}\langle E_{i} \rangle^{2}}) - 1) + \frac{K}{2}( \frac{J^{2} - 1}{2} - \ln{J} ),
\end{align}
where $\langle \cdot \rangle$ denotes the Macaulay bracket. In this model, the fiber strain of the two fiber groups is expressed as:
\begin{equation}\label{eqn:fiberstrain}
    E_{i} = \kappa (I_{1} - 3) + (1 - 3\kappa)(I_{4i} - 1),\quad i=1,2,
\end{equation}
where $k_{1}$ and $k_{2}$ are fiber modulus and the exponential coefficient, respectively, $I_{1}$ is the first principal invariant, and $I_{4i}$ is the fourth principal invariants corresponding to the $i-$th fiber group. Mathematically, for the $i-$th fiber group with angle direction $\alpha_i$ from the reference direction, $I_{4i}$ is calculated by $\bm{n}_i^\text{T}\mathbf{C}\bm{n}_i$, where $\mathbf{C}$ is the right Cauchy-Green tensor and $\bm{n}_i=[\text{cos}\alpha_{i}, \text{sin}\alpha_{i}]^\text{T}$. In our simulations, we consider a material with fiber reinforcement in the vertical direction (see Fig.~\ref{fig:hyper_schematic} right for illustration). Therefore, for both fiber groups we set $\alpha_{i}=\pi/2$. In Eq.~\eqref{eqn:fiberstrain}, fiber dispersion is denoted as $\kappa$, whose value ranges from $0$ to $\frac{1}{3}$. Intuitively, $\kappa = 0$ means no fiber dispersion whereas $\kappa = \frac{1}{3}$ represents an isotropic fiber dispersion. In this example, we consider the fiber oriented vertically with no dispersion ($\kappa = 0$). All parameter values in this example are summarized in Table~\ref{table:hyper_param}.

We now present the problem set up of this example. As depicted in Fig.~\ref{fig:hyper_schematic}(a), we consider a 2D unit square plate with a centered circular void (radius as $0.1$). The plate deforms under a uniaxial tension, $\Tb_0(\xb)$, applied on its top edge. The bottom edge is clamped. We model the material response of the entire domain using an SPH model, whose solution is taken as the ground-truth solution. More details of the SPH model are provided in~\ref{app:sph}. To develop a coupling model, we consider a similar setting as in the previous example and decompose the entire domain $\omg$ into two non-overlapping subdomains. Then, we train DeepONets using SPH data to obtain a surrogate for the internal domain while the external region is described by a continuum FEM model. Models in these two domains communicate by exchanging proper interface conditions on their common interface, $\Gamma$. 
In Fig.~\ref{fig:hyper_schematic}(b) we present a schematic of information transmission of the coupling framework with a N-D method. At the $n$-th interation, the FEM solver receives an updated distributed traction on $\Gamma$ from DeepONets and solves the updated displacement field, $\ub^n_{FEM}$, with the given information. Then, the FEM transmits the updated displacement information $\ub^n_{FEM}|_{\Gamma}$ to DeepONets. With the displacement on $\Gamma$ as input, DeepONets estimate the first Piola-Kirchhoff (PK1) stresses in $\Omega_{NN}$. Based on the predicted PK1 stresses, we then calculate the surface traction on $\Gamma$ and other associated quantities, such as the equivalent plastic strain $\bar{\varepsilon}^\text{p}$ and von Mises stress $\bar{\sigma}$, accordingly. At the $n$-th iteration, the system solution at the interface $\Gamma$ is updated as $\Tilde{\Tb}(\xb) = (1-\theta)\Tb^{n}_{I}(\xb) + \theta \Tb^{n}_{II}(\xb)$.

Next, we briefly describe the training process of DeepONet. To generate the training/testing dataset, we sample $1,000$ different traction loading $\Tb_0(\xb)$ on the top edge from a random field (see Algorithm in Sec.~\ref{sec:rand_field}). Then, for each sampled traction loading, we perform an SPH simulation to obtain the solutions in the entire domain and collect the corresponding solutions of displacement and PK1 stress fields of in $\Omega_{NN}$. Among these 1000 samples, 900 cases are employed as the training data while the rest is kept as testing data. As depicted in Fig.~\ref{fig:hyper_schematic}(c), we consider two approaches for network training. In the first approach, the network ($Net_{u2s}$) takes the displacement at the interface ($\ub|_{\Gamma}$) as input and predicts PK1 stress as the output. In the second approach, the interfacial traction $T|_{\Gamma}$ is employed as the input of the network ($Net_{s2u}$), yielding the displacement field as output. These approaches provide a flexibility of imposing different interface conditions. We adopt the first approach in the N-D method and combine the information of the two approaches ($\hat{\ub}$ and $\hat{\Tb}$) in the R-D/R-N method.

\begin{figure}
    \centering
	\includegraphics[width=1.0\textwidth]{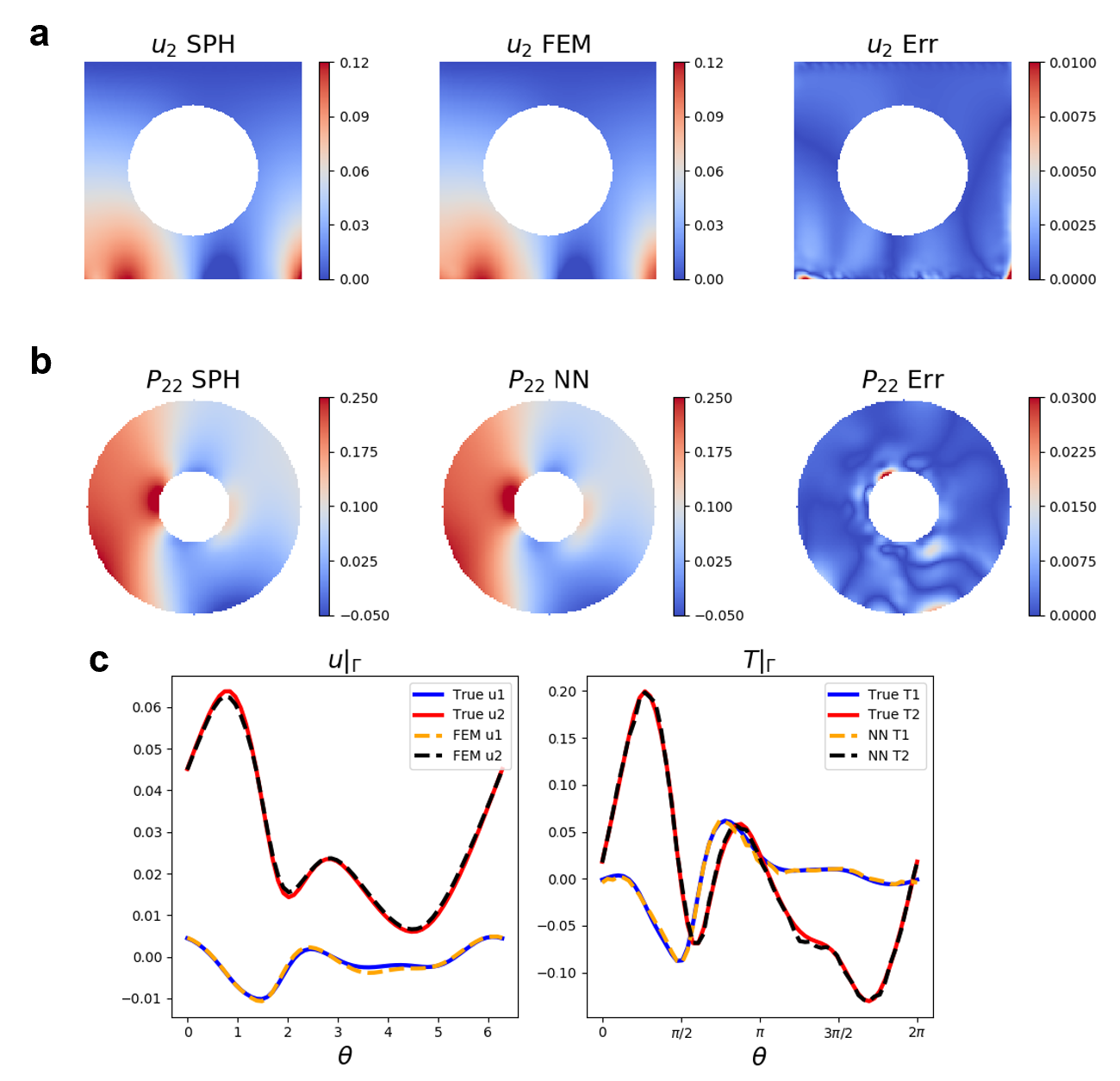} 
    \caption{\textbf{Results of coupling FEM and DeepONet in the hyperelasticity problem.} Results of (a) displacement in $\Omega_{FEM}$ and (b) $P_{22}$ in $\Omega_{NN}$. From left to right: prediction from SPH, prediction from FEM, and their absolute differences. (c) Predicted displacement and traction at the interface. Predictions from both FEM and DeepONets are compared with the true data. Solid lines: displacement and traction of true data in x (blue) and y (red) directions. Dashed lines: displacement and traction of the model predictions in x (yellow) and y (black) directions.}
    \label{fig:hyperelastic}
\end{figure}

Fig.~\ref{fig:hyperelastic} shows the coupling results of a typical testing case with a N-D method. In Fig.~\ref{fig:hyperelastic}(a-b), we compare the vertical displacement $u_2$ in $\Omega_{FEM}$ (first row) and the PK1 stress component $P_{22}$ in $\Omega_{NN}$ (second row) between the SPH ground truth (first column) and the FEM/NN prediction (second column). The prediction errors are displayed in the third column. We observe that the coupling results match well with the ground truth solution with the largest prediction errors distributed near the bottom corners. The errors are partially induced by the numerical interpolations in the FEM and reduced regularity in that region. To further examine the prediction accuracy, in Fig.~\ref{fig:hyperelastic}(c) we plot a quantitative comparison of displacement and traction on the interface $\Gamma$ as functions of the polar coordinate angle $\theta$. Despite some numerical discrepancies between SPH and FEM (see~\ref{app_sec:sph}), the FEM/DeepONet predictions shown in dashed lines successfully reproduce the SPH simulation results depicted in solid lines. Therefore, our proposed method is capable of capturing the mechanics from SPH with a substantially improved efficiency: the time cost of performing an SPH simulation is approximately 4 hours whereas running its surrogate just takes a fraction of a second (Table~\ref{table:walltime}). Admittedly, the excessive cost of SPH is exacerbated because we use a time-dependent SPH solver to simulate a static problem. Nonetheless, we can see that replacing a particle model with its surrogate in a multiscale coupling framework poses unique advantages in both efficiency and programming easiness.

\begin{table}
\centering
\begin{adjustbox}{}
 \begin{tabular}{|c | c |} 
 \hline
 Model & Wall Time \\
 \hline
 FEM-SPH & $ \thicksim 4h$ \\
 \hline 
 FEM-DeepONet & $<1s$ \\
 \hline
\end{tabular}
\end{adjustbox}
\caption{Wall time comparison for SPH and DeepONet per iteration in the hyperelastic problem. The excessive cost of SPH is exacerbated because we use a time-dependent SPH solver to simulate a static problem.}
\label{table:walltime}
\end{table}

\begin{figure}
    \centering
	\includegraphics[width=0.5\textwidth]{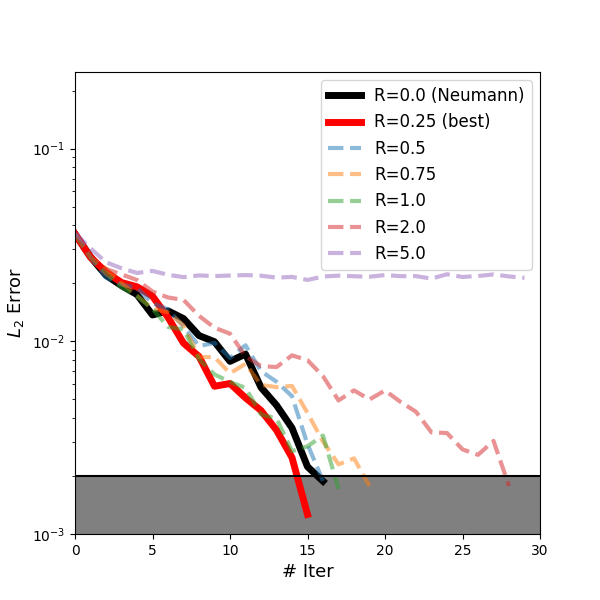} 
    \caption{\textbf{$L_{2}$ errors of the interfacial displacement in the hyperelastic multiscale model with a Robin boundary.} The red line with Robin coefficient R=0.25 converges fastest among the testing cases. The black solid line represents the relative error history of a Neumann boundary condition (R=0). The shaded area indicates that the relative error is less than 2\%. Notice that we present the error with respect to the ground truth to show the convergence and accuracy of our framework.}
    \label{fig:sph_robin_converg}
\end{figure}

To further illustrate the flexibility and investigate the convergence rate of our coupling framework, we employ a R-D/R-N coupling method to our framework with a variety of values of the Robin coefficient $R$. To clarify, R-D/R-N means that the FEM, imposed with a Robin-type boundary condition, separately transmits the information of displacement and traction to $Net_{u2s}$ and $Net_{s2u}$ in order to update the Robin information in the next step. The continuum model with a Robin boundary condition is modified as:
\begin{align}
    \Psi & = \int_{\Omega}\psi(\ub)\text{d}\xb - \int_{\Omega}\bb\cdot \ub \text{d}\xb - \int_{\Gamma_{N}} \Tb_0\cdot \ub \text{d}s -\int_{\Gamma_{R}} \left(\rb-\frac{1}{2}R \ub\right)\cdot \ub \text{d}s.
\end{align}
$\rb$ is the Robin boundary condition applied on the interface $\Gamma_R$. We define $\rb$ as:
\begin{equation}\label{eqn:robin}
    \rb = \hat{\Tb} + R\cdot \hat{\ub}, \text{ for } \xb \in \Gamma_{R}, 
\end{equation}
where $R$ is the (positive) Robin coefficient, $\hat{\Tb}$ and $\hat{\ub}$ are the known traction and displacement. 

The coupling procedure is changed as well. At the $n$-th iteration, we first solve the FEM model with a Robin boundary condition $\Tilde{\rb}^{n}(\xb)$. Then, we transmit the interfacial traction and displacement of FEM to $Net_{u2s}$ and $Net_{s2u}$, respectively. The corresponding displacement $\hat{\ub}^{n+1}$ and tractions $\hat{\Tb}^{n+1}$ are estimated by DeepONets and used to update the Robin boundary condition: $\Tilde{\rb}^{n+1}(\xb) = (1-\theta)\rb^{n+1}_{FEM}(\xb) + \theta \rb^{n+1}_{NN}(\xb)$. Here, $\theta=0.5$ is taken as a fixed relaxation parameter.

We further study the convergence of interfacial displacement errors with different values of $R$. In Fig.~\ref{fig:sph_robin_converg}, we highlight the results from the N-D coupling method with the black line and the results of R-D/R-N with $R=0.25$ (best result) in red. The shaded area indicates 2\% equivalent relative error of the test case. Due to the nonlinearity of the problem, it generally takes more iterations for the coupling framework to reach the stopping criterion. When taking a R-N coupling method with a sub-optimal Robin coefficient (such as $R=5.0$ as shown in the purple line of Fig.~\ref{fig:sph_robin_converg}), the coupling framework fails to converge. This fact again demonstrates the importance of choosing an appropriate coupling method. 

\section{Discussion} 
\label{sec:discussion}

In this paper, we propose an efficient concurrent coupling framework for multiscale modeling of mechanics problems. In lieu of coupling an expensive microscopic model, we propose to employ a deep neural operator, DeepONet, as a surrogate to approximate the microscopic solution in the domain with fine-scale features. The response in the coarse-scale domain is simulated by a standard numerical model, such as the finite element method. The two models are coupled concurrently by exchanging information at the interface until convergence. To verify the performance of this framework, we study four benchmarks including solving static and dynamic problems for different materials. We have also demonstrated that the framework is readily applicable for various interface conditions. The results show that the cases with a Robin boundary condition tend to converge faster than a Neumann/Dirichlet boundary condition. Moreover, the predictions of the coupled model agree well with the true solution acquired from a numerical method, indicating the generalization ability of DeepONet and the accuracy of the proposed framework. 


In addition, using a neural operator as surrogate enables the model to be trained directly from data. Such a model is particularly promising for learning dynamics of complex materials without explicit constitutive models. 
Moreover, coupling DeepONet with FEM substantially improves the computational efficiency in multiscale modeling: DeepONet predicts the expensive microscopic behavior only at a fraction of second. Hence, the overall computational cost of the proposed framework could be shorten by orders of magnitude than the existing multiscale coupling methods. In addition, we can train the surrogate model to learn based on partial information: it can predict the information only at the interface and neglect dynamics inside the microscopic region. Thus, the overall framework behaves as an artificial-intelligence type boundary condition, which would potentially improve the efficiency especially in the scenarios where only the macroscopic dynamics is of interest.

As a further note, we would like to point out the keys to build a successful neural operator based coupling framework with guaranteed convergence and to avoid possible pitfalls. 
First, the generalization ability of DeepONet determines if the result converges and its convergence rate. Utilizing an accurate neural operator model takes fewer iterations to converge. However, a poorly trained model could lead to slow convergence or even divergence. Second, normalizing data is another key to convergence. In our experiments, the training data are at disparate scales, ranging from $10^{-3}-10^{1}$. Properly normalization of the training data would be essential to the convergence of numerical iterations. In addition, choosing proper coupling strategies, such as the right Robin coefficient, also plays a critical role in guaranteeing fast numerical convergence as it affects the condition number of the stiffness matrix in FEM and the coupling system~\cite{dijkstra2006condition}. A large condition number may lead to a slow convergence or even divergent result in the coupling framework (see Sec.~\ref{sec:hyper} and~\cite{gustafson1998domain,douglas1997accelerated}). 

Certainly, more improvements and directions can be considered in the future. Learning directly from data facilitates a unique feature that DeepONet can learn data that comes from different scales, which has been demonstrated in Sec.~\ref{sec:hyper}. In the future, it would be interesting to develop a data-driven model from noisy data, such as molecular dynamics and dissipative particle dynamics, as presented in Fig.~\ref{fig:schmematic}. In addition, overcoming the multiscale characteristic length and time scale would be another challenge in multiscale modeling with machine learning, that is, the time and length of a microscopic model are usually orders of magnitude smaller than the continuum model, causing challenges in network training and long-term predictions. The coupling results may drift away from the true solution or even diverge due to inaccurate predictions from surrogate models. Another improvement would be considering microstructures and geometric variations (see~\cite{masi2021thermodynamics}). Developing an operator-learning neural network that can predict dynamics with different geometric variations would be a great improvement for broadening the application of the proposed method. Also, simulating fracture progression is another natural and promising application of the multiscale coupling framework. As fracture progresses, the field of interest that includes the damage region may also move along with the tip of a crack, posing a challenge in both coupling algorithm and network training. Our framework can be further extended to employ other coupling methods such as quasicontinuum~\cite{tadmor1996quasicontinuum}, multigrid~\cite{brandt1977multi}, heterogeneous multiscale method (HMM)~\cite{weinan2003heterognous}, etc. 


\section*{Acknowledgment}
MY, EZ, and GEK acknowledge the support by grant U01 HL142518 from the National Institutes of Health. Y. Yu would like to acknowledge support by the National Science Foundation under award DMS 1753031.

\appendix
\section{Smoothed Particle Hydrodynamics}
\label{app:sph}
In this section, we briefly introduce the basic formulation of the Total Lagrangian Smoothed Particle Hydrodynamics (TLSPH). We refer the reader to~\cite{rausch2017modeling,ganzenmuller2015hourglass} for more details. In the SPH framework, physical quantities are approximated with the neighboring information in a kernel. Consider a function $f(X)$ at $X_{i}$ can be approximated by $\hat{g}(X_{i})$ with the integration
\begin{equation}
    \hat{g}(X_{i}) = \int f(X)W(X-X_{i}) \text{d}X,
\end{equation}
where $W(X)$ is a weighting kernel which is chosen as a third-order polynomial
\begin{equation}
  W(R_{j}, h) = A
    \begin{cases}
      (h-R_{j})^{3}, & R_{j} < h, \\
      0, & R_{j} \leq h,
    \end{cases}       
\end{equation}
$h$ denotes the radius of the integration kernel; $R_{j}$ is defined as the distance between $X_{j}$ and the reference point in the kernel $X_{i}$ with $A = 10 / \pi h^{5}$ in two dimensions or $A = 10/\pi h^{6}$ in three dimensions. SPH numerically approximates the integration as
\begin{equation}
    g(X_{i}) = \sum_{j\in S} f_{j} V_{j} W(R_{j}, h),
\end{equation}
where $V_{j}$ is the volume of particle $j$. The gradient of $g(X_{i})$ with respect to its reference coordinate is
\begin{equation}
    \nabla_{X}g(X_{i}) = \sum_{j \in S} f_{j} V_{j} \nabla_{X}W(R_{j}, h)
\end{equation}
with
\begin{equation}
    \nabla_{X}W(R_{j}, h) = (\frac{\partial W(R_{j}, h)}{\partial R_{j}}) \frac{R_{j}}{R_{j}}
\end{equation}
For a vector $f(X)$, its gradient to the referential coordinate is approximated as the following using the integration rule:
\begin{equation}
    \nabla_{X} g(X_{i}) = \sum_{j\in S} f_{j} \otimes V_{j} \nabla_{X}W(R_{j}, h)
\end{equation}

We then introduce two \textit{ad-hoc} corrections for keeping symmetrization and first-order completeness~\cite{monaghan1988introduction}.
\begin{equation}
    \nabla_{X}g(X_{i}) = \sum_{j \in S} (f_{j} - f_{i}) V_{j} \nabla_{X}W(R_{j}, h).
\end{equation}
Another correction guarantees the first-order completeness~\cite{randles1996smoothed}, the corrected gradient tensor is defined as
\begin{align}
    \Tilde{\nabla}_{X}W(R_{j}, h) = A^{-1}\nabla_{X}W(R_{j}, h)
\end{align}
where the shape tensor $A$ is
\begin{equation}
    A_{i} = \sum_{j \in S} V_{j} \nabla_{X} W(R_{j}, h) \otimes R_{j}
\end{equation}

We introduce the SPH integration to the governing equation of solid at the continuum level. The equilibrium equation is:
\begin{equation}
    \nabla \cdot P + \rho_{0} \bb = \rho_{0}\ddot{x}
\end{equation}
where $P$ is the first Piola-Kirchhoff stress tensor, $\bb$ the body force vector, $\rho_{0}$ the referential mass density, and $\ddot{x}$ the acceleration. We define the deformation gradient as
\begin{equation}
    F_{i} = \frac{\partial x_{i}}{\partial X_{i}} = \sum_{j \in S} r_{j} \otimes V_{j} \Tilde{\nabla}_{X}W(R_{j}, h)
\end{equation}
According to basic continuum mechanics law, 
\begin{equation}
    P_{i} = F_{i} S_{i} = 2 F_{i} \frac{\partial \mathcal{W} }{ \partial C }
\end{equation}
As proved in~\cite{bonet1999variational}, the internal forces emerge from the divergence of the first Piola-Kirchhoff stress $P$:
\begin{equation}
    m_{i}\ddot{x}_{i} = f^{int} + f^{ext},
\end{equation}
where the internal forces is expressed as
\begin{equation}
    f_{i}^{int} = \sum_{j \in S} V_{i} V_{j} (P_{i} \Tilde{\nabla}_{X}W(R_{i}, h) - P_{j}\Tilde{\nabla}_{X}W(R_{i}, h) ).
\end{equation}

We adopt another correction for suppressing spurious hourglassing mode~\cite{ganzenmuller2015hourglass}
\begin{equation}
    f_{i}^{hg} = \sum_{j \in S} -\alpha \frac{E V_{i} V_{j} W(R_{j}, h) }{2 R_{j}^{2}}(\delta_{i} + \delta_{j})\frac{r_{j}}{r_{j}}
\end{equation}
The deformation derivative in time is approximated as
\begin{equation}
    \dot{F} = \frac{1}{\Delta t}(F^{t+1} - F^{t})
\end{equation}

We adopt the strain energy function $\mathcal{W}(\textbf{C}, \textbf{M})$ for a fiber-enhanced tissue proposed in~\cite{holzapfel2000new}
\begin{equation}
    \mathcal{W}(C, M) = \frac{c}{2}(I_{1} - 3) - c\ln(J) + \frac{k_{1}}{2k_{2}}\sum^{2}_{i=1}(\exp{(k_{2}\langle E_{i} \rangle^{2}}) - 1) + \frac{K_{0}}{2}( \frac{J^{2} - 1}{2} - \ln{J} ),
\end{equation}
where principal invariants, $I_{1}$ and $I_{4}$ are defined as,
\begin{equation}
    I_{1} = C:I, J = \det F, I_{4} = C:M\otimes M
\end{equation}
The parameters are set the same as the FEM model in Sec.~\ref{sec:hyper}. We refer the reader to \cite{rausch2017modeling,ahmadzadeh2019modeling} for more details.

\section{Accuracy of FEM and SPH}




\subsection{Accuracy of SPH}
\label{app_sec:sph}

\begin{figure}
    \centering
	\includegraphics[width=1.0\textwidth]{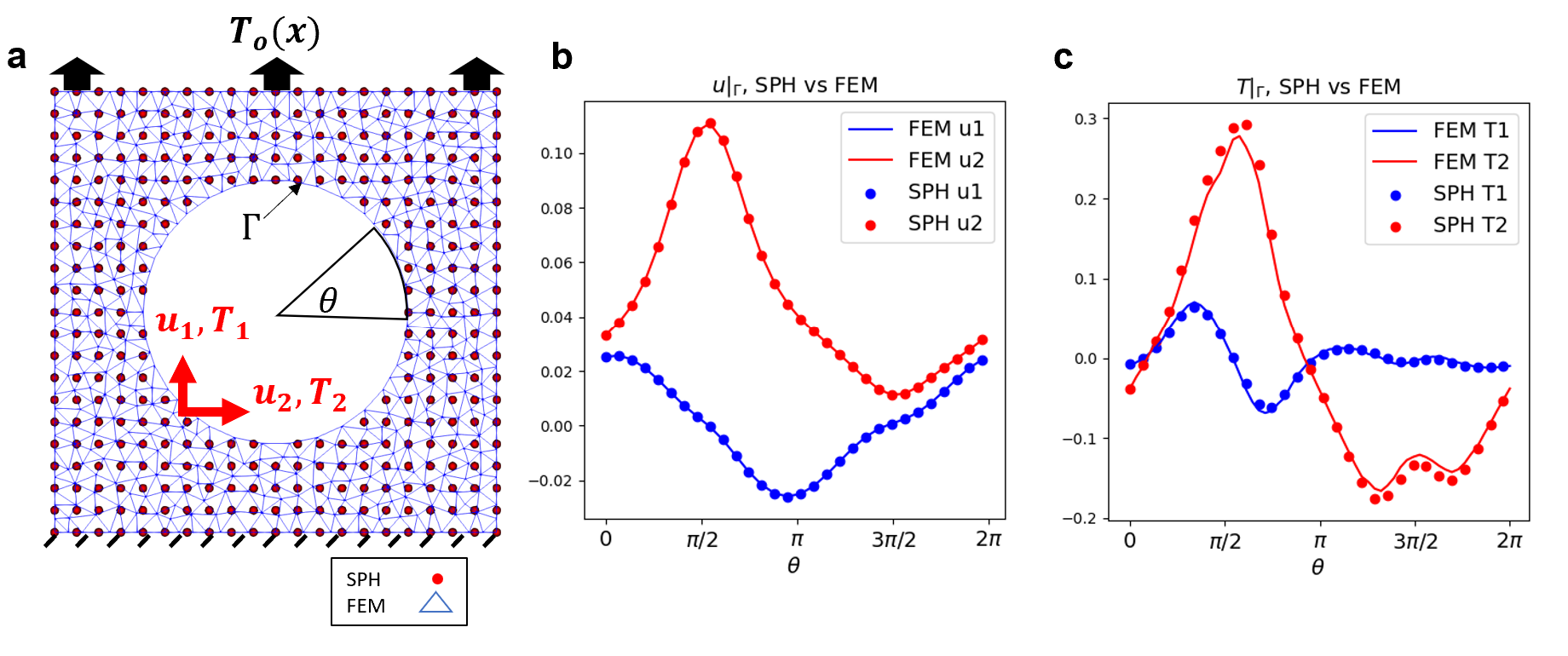} 
    \caption{\textbf{Comparison of SPH and FEM.} (a) We impose the same Dirichlet boundary condition on the top and internal circle for a FEM and SPH model. (b-c) Interfacial displacement and traction of both models.}
    \label{fig:append_SPH_FEM}
\end{figure}

To compare the numerical solution of the FEM and SPH model, We present their computational results for a same benchmark problem. The problem is set up as shown in Fig.~\ref{fig:append_SPH_FEM}(a). We impose a distributed traction boundary condition on the top edge and impose a displacement boundary condition (Dirichlet-type) at the interface for both models shown in Fig.~\ref{fig:append_SPH_FEM}(b): the displacement in $x$ and $y$ at the interface are plotted against the angle $\theta$. The corresponding tractions at the interface are presented in Fig.~\ref{fig:append_SPH_FEM}(c). The solid lines are the FEM tractions in $x$ and $y$ directions with the dashed lines denoting the SPH results. The results correspond well in general with some deviation around $\theta=\pi/2$ and $\theta=3\pi/2$ due to the difference in numerical integration methods. This discrepancy partially contributes to the overall errors observed in Sec.~\ref{sec:hyper}.

\section{Network Training Details}
\label{sec:net_train}

\begin{figure}
    \centering
	\includegraphics[width=1.0\textwidth]{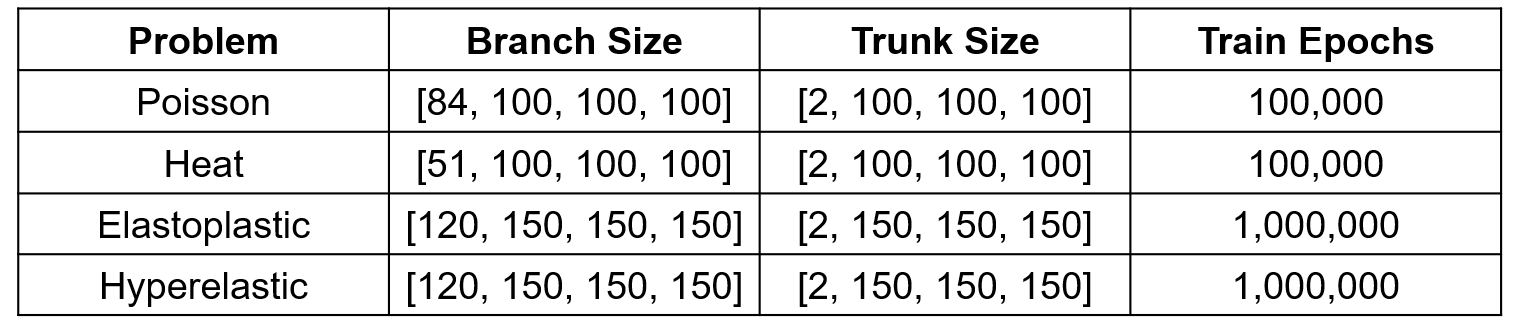}
    \caption{\textbf{Network size for various problems.} We tabulate the branch/trunk network size for the four examples in Sec.~\ref{sec:results} with the number of training epochs.}
    \label{fig:net_size}
\end{figure}

\begin{figure}
    \centering
	\includegraphics[width=0.7\textwidth]{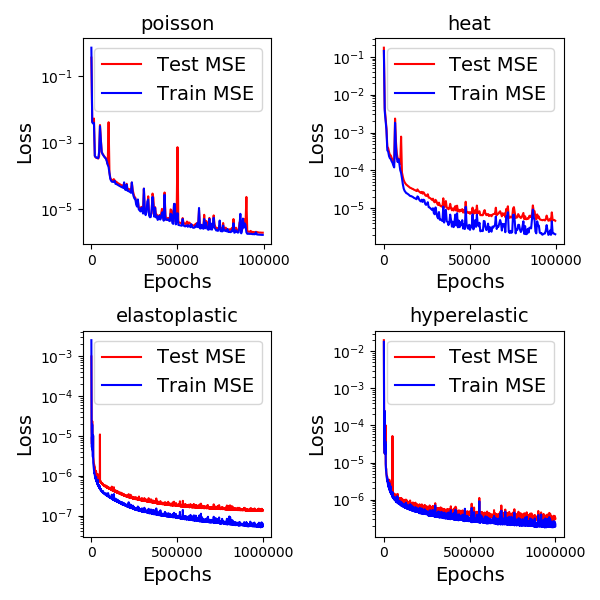} 
    \caption{\textbf{History of Training/testing errors of DeepONet for different problems.}}
    \label{fig:net_train}
\end{figure}

\begin{figure}
    \centering
	\includegraphics[width=1.0\textwidth]{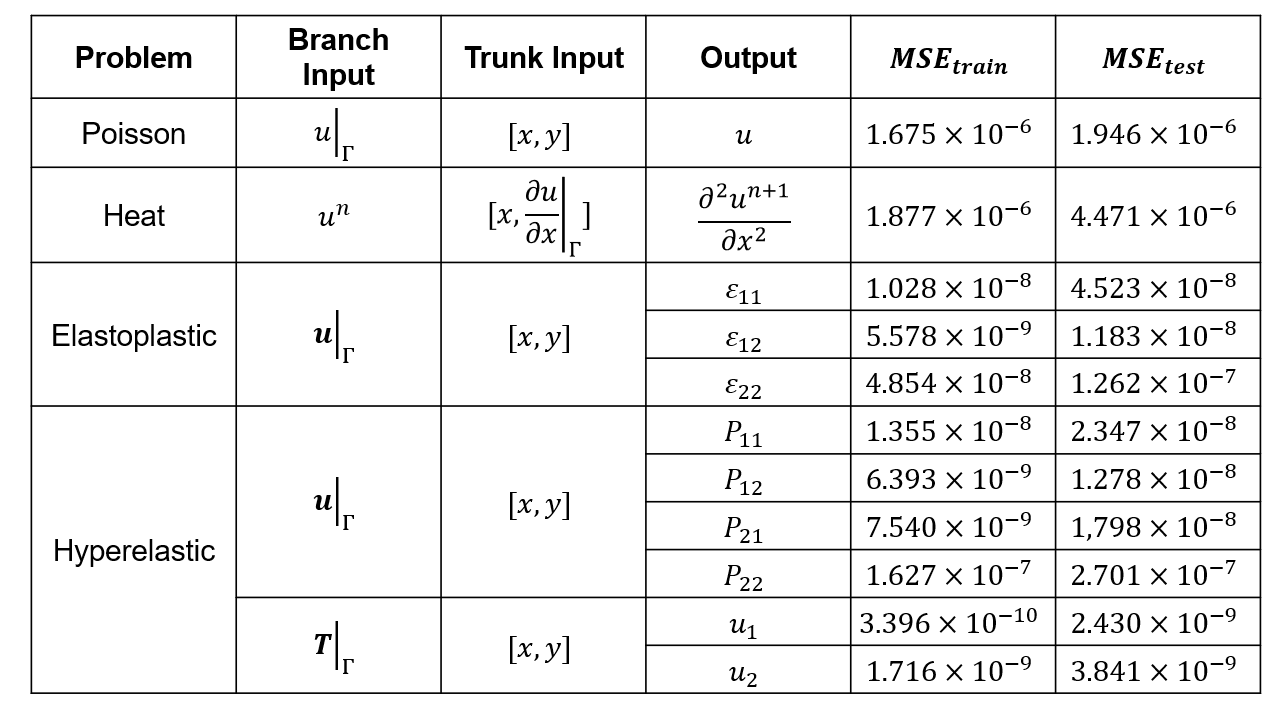} 
    \caption{\textbf{Details of network setup (inputs/outputs and MSE errors of training/testing) for each example in Sec.~\ref{sec:results}.}}
    \label{fig:io_net}
\end{figure}

We explain the details of network training in this section. Network architecture for each problem is presented in Fig.~\ref{fig:net_size}. We adopt a fully-connected neural network (FNN) with four layers for all the sub-networks. The input layer dimension of the branch net depends on the number of points at the interface, which ranges from 51 to 120 in our examples. Three hidden layers are concatenated with the input layer where we adopt the hyperbolic tangent function as the activate function. Each hidden layer has 100 or 150 neurons shown in Fig.~\ref{fig:net_size}. To minimize the loss function for each case, we set the training epoch as 100,000 steps for the heat and Poisson example and 1,000,000 for the other two examples. Moreover, we illustrate the error history of a few cases in Fig.~\ref{fig:net_train} where each cases shows a relatively small difference between the training and testing errors. The low testing errors indicating a good generalization capability of the network. More specifically, we tabulate the input/output of each network in Fig.~\ref{fig:io_net} with MSEs of the training and testing data for each case.

\section{Data Generation}
\label{sec:rand_field}

\subsection{Random field generation}
We provide an overview of the algorithms of random fields generation using Fast Fourier Transformation (FFT). Let $W(\xb)$ be a Gaussian white noise random field on $\mathbb{R}^{d}$. A random field $\phi(\xb)$ can be sampled by
\begin{equation}
\label{equ:grf}
    \phi(\xb) = \mathcal{F}^{-1}(\gamma^{1/2}\mathcal{F}(W))(\xb)
\end{equation}
where $\mathcal{F}$ and $\mathcal{F}^{-1}$ denote Fourier transformation and its inverse. $\gamma$ represents a correlation function $\norm {k}^{-\alpha}$ where $\norm k$ is the $L_{2}$ norm of the wave number $k \in \mathbb{R}^{d}$. In Sec.~\ref{sec:poisson}, \ref{sec:heat}, and \ref{sec:hyper}, we adopt this method for generating the training data with $\alpha=5$. We refer the reader to find more theoretical details in~\cite{LangPotthoff2011}


\subsection{Sampling in elastoplasticity}
\label{subsec:rand_ep}

In the case of elastoplasticity, the non-uniform tension $t_0(x)$ (see Fig.~\ref{fig:elastoplasticity}) is generated by
\begin{equation}
\label{eqn:plasticity_Fourier}
    t_0(x) = \sum_{i=1}^{3}\frac{A_i}{i}\cos{(ix)}+\sum_{i=1}^{3}\frac{B_i}{i}\sin{(ix)},
\end{equation}
where $A_i,B_i\sim N(0,0.05^2)$ ($i\in\{1,2,3\}$) are independent normal random variables. We choose the standard deviation as $0.05$ to make sure that the dataset contains similar numbers of cases with and without plastic deformation.


\bibliographystyle{abme}
\bibliography{reference}
\end{document}